\documentclass[lettersize,journal]{IEEEtran}
\usepackage{amsmath,amsfonts}
\usepackage{algorithm}
\usepackage{array}
\usepackage{textcomp}
\usepackage{stfloats}
\usepackage{url}
\usepackage{verbatim}
\usepackage{graphicx}
\usepackage{cite}
\usepackage{xcolor}
\usepackage{dsfont}
\usepackage{algpseudocode}
\usepackage{lineno,hyperref}
\usepackage{array}
\usepackage{multirow}
\usepackage{subfigure}
\usepackage{amsmath}
\usepackage{epstopdf}
\usepackage{framed} 
\usepackage{booktabs}
\usepackage{enumitem}
\begin{document}

\title{\textcolor{black}{Augmented Physics-Based Li-ion Battery Model via Adaptive Ensemble Sparse Learning and Conformal Prediction}}

\author{Samuel Filgueira da Silva, Mehmet Fatih Ozkan, Faissal El Idrissi and Marcello Canova
        % <-this % stops a space
\thanks{Department of Mechanical and Aerospace Engineering, Center for Automotive Research, The Ohio State University, Columbus, OH, USA. Emails: 
        {\tt\small filgueiradasilva.1@osu.edu, ozkan.25@osu.edu,  elidrissi.2@osu.edu, canova.1@osu.edu}}% <-this % stops a space
\thanks{}}

\markboth{Preprint submitted to IEEE Transactions on Transportation Electrification}%
{Author \MakeLowercase{\textit{et al.}}: [Short Paper Title]}

\maketitle
\begin{abstract}
Accurate electrochemical models are essential for the safe and efficient operation of lithium-ion batteries in real-world applications such as electrified vehicles and grid storage. Reduced-order models (ROM) offer a balance between fidelity and computational efficiency but often struggle to capture complex and nonlinear behaviors, such as the dynamics in the cell voltage response under high C-rate conditions. To address these limitations, this study proposes an Adaptive Ensemble Sparse Identification (AESI) framework that enhances the accuracy of reduced-order li-ion battery models by compensating for unpredictable dynamics. The approach integrates an Extended Single Particle Model (ESPM) with an evolutionary ensemble sparse learning strategy to construct a robust hybrid model. In addition, the AESI framework incorporates a conformal prediction method to provide theoretically guaranteed uncertainty quantification for voltage error dynamics, thereby improving the reliability of the model’s predictions. Evaluation across diverse operating conditions shows that the hybrid model (ESPM + AESI) improves the voltage prediction accuracy, achieving mean squared error reductions of up to 46\% on unseen data. Prediction reliability is further supported by conformal prediction, yielding statistically valid prediction intervals with coverage ratios of 96.85\% and 97.41\% for the ensemble models based on bagging and stability selection, respectively.
\end{abstract}


\section{Introduction}

\IEEEPARstart{A}{ccurate} prediction of lithium-ion battery (LiB) cell dynamics is crucial for optimizing performance, extending life, and ensuring safety across diverse applications, from consumer electronics to electric vehicles and grid storage \cite{cano2018batteries,CHEN20194363}. Achieving this requires modeling approaches that can reliably capture the LiB response. Physics-based models (PBMs) offer a rigorous, interpretable, and generalizable framework for capturing internal dynamics and their effects on terminal voltage \cite{ALI2024144360}. By incorporating fundamental electrochemical processes, high-fidelity models provide a detailed representation of LiB behavior through coupled partial differential equations (PDEs) that describe key physical mechanisms such as interfacial reaction kinetics, mass and charge transport, and equilibrium potentials. However, their high computational cost makes them unsuitable for real-time battery management applications \cite{YU2023107661}. 

Reduced-order models (ROMs) reduce computational burden but often sacrifice accuracy, particularly at high C-rate conditions \cite{fan2016modeling}. This issue stems from the inherent simplifications, such as idealized electrode structures, uniform reaction kinetics, and constant material properties, which neglect complex, state-dependent dynamics. As a result, these modeling gaps can introduce calibration errors, as parameter identification must compensate for unmodeled behavior. Umeasurable internal states (e.g., lithium concentration) further contribute to estimation uncertainty \cite{lai2021new}, requiring complex identification techniques such as sensitivity analysis, parameter grouping, and optimal experiment design \cite{huang2023reinforcement,mikesell2025real}.

To address these limitations, machine learning (ML) models have gained traction in LiB research. Deep learning has been used to approximate input/output behavior \cite{liang2022comparative,miranda2023particle}, offering efficiency gains but lacking interpretability, generalizability and incorporation of latent electrochemical states. These limitations motivate hybrid models that combine physics-based structure with data-driven flexibility \cite{amiri2024lithium}. Recent work has explored hybrid frameworks that integrate physics-based models with ML components to capture unmodeled dynamics \cite{pozzato2023combining}. Artificial neural networks (ANNs) have shown improvements in voltage prediction \cite{tu2023integrating}, but function as black-boxes, limiting physical insight and parsimony. Moreover, most methods apply a static correction to the voltage error without modeling its temporal evolution. Capturing the dynamics of voltage error enables modeling of cumulative effects like diffusion delays and hysteresis, which are prevalent in LiBs \cite{krewer2018dynamic}. %To address these limitations, hybrid modeling techniques that enforce physical consistency and enhance interpretability are especially promising, offering the potential to model temporal dependencies while maintaining transparency. %Therefore, hybrid learning techniques that emphasize transparency and physical consistency are particularly promising, as they can improve model interpretability.

%% introduce physics-inspired ML approaches: show that they have been successfully applied to different engineering applications

Scientific machine learning techniques present a promising approach for addressing these requirements, as they have been effectively leveraged across diverse fields, including physics \cite{de2020discovery}, biology \cite{zhang2024discovering}, and chemistry \cite{bhatt2023sindy}, enabling enhanced modeling, prediction, and discovery of complex system behaviors. Methods such as Sparse Identification of Nonlinear Dynamics (SINDy) \cite{AHMADZADEH2024110743}, Koopman operator \cite{choi2023data}, and Symbolic Regression (SR) \cite{flores2022learning}, have been explored for learning the governing equations of electrochemical battery models directly from data measurements. However, SINDy and Koopman rely on predefined function libraries, which require prior knowledge of the system dynamics and may not fully capture complex nonlinearities. SR can become computationally intractable in high dimensions \cite{keren2023computational}. 

A recent study \cite{da2024improving} introduced a hybrid sparse system identification method that jointly tunes function libraries and hyperparameters to correct voltage predictions of a reference model. While effective, the method does not tackle the challenges of limited data, a common occurrence in battery testing. Here, ensemble learning becomes valuable by training on multiple data subsets, enhancing robustness and generalization, particularly in sparse data conditions \cite{sashidhar2022bagging}.

\vspace{.01mm} %5mm vertical space
While ensemble techniques significantly enhance robustness and predictive stability, another critical aspect involves quantifying model uncertainty and assessing prediction reliability, particularly when extrapolating beyond training data. Gaussian Process Regression (GPR) has been widely used for uncertainty estimation in state of charge (SOC) \cite{lee2023battery} and state of health (SOH) \cite{yang2018novel} prediction. However, GPR assumes Gaussian error distribution, which may not hold in real battery systems. This assumption can yield overconfident or misleading uncertainty estimates under real-world conditions. 
\vspace{.01mm} %5mm vertical space

In contrast, Conformal Prediction (CP) provides a model-agnostic, distribution-free framework for generating prediction intervals with theoretical validity \cite{vovkconformal, manokhin2023practical}. CP can be applied on top of any point prediction model, including neural networks, decision trees, or ensemble methods. It has gained significant traction in time-series prediction models due to its ability to quantify predictive uncertainty without relying on strong parametric assumptions \cite{manokhin2023practical}. %This property makes CP particularly promising for safety-critical applications in battery management systems, where reliable uncertainty quantification is as crucial as the accuracy of the predictions themselves \cite{CP_Sachan}. %Conformal prediction has been used in different real-world applications including vehicle trajectory  due to its generality, simplicity, and theoretical guarantees.
\vspace{.05mm} %5mm vertical space

In light of the current state of the art, a clear research gap remains in developing robust sparse learning methods for hybrid battery modeling that also provide rigorous uncertainty quantification. This work addresses that gap by proposing a framework to identify and correct the voltage error dynamics between a reduced-order LiB model and experimental measurements, while simultaneously generating statistically valid prediction intervals for those corrections.
\vspace{.05mm} %5mm vertical space

The study introduces an evolutionary adaptive sparse identification method that captures the unmodeled dynamics of the battery by augmenting the predictions of a physics-based model with a data-driven correction. This is achieved through ensemble learning strategies that enhance model robustness and generalization. Additionally, the framework incorporates conformal prediction to quantify model uncertainty, enabling the construction of prediction intervals with theoretical coverage guarantees across a range of operating conditions. This improves the reliability of the voltage predictions and supports their deployment in safety-critical applications. To the best of the author's knowledge, this is the first study to combine adaptive sparse identification with ensemble learning and conformal prediction for hybrid modeling of LiBs.

%\begin{itemize}
%    \item  An evolutionary adaptive sparse identification approach is proposed to model the missing dynamics in lithium-ion battery models. By incorporating ensemble learning, the method aims to capture the voltage correction of the physics-based model.

%    \item  Model uncertainty is systematically quantified using conformal prediction, providing statistically valid prediction intervals for the hybrid battery model across various operating conditions, thereby improving reliability in the voltage error predictions.
%\end{itemize}

The rest of this manuscript is organized as follows: Section \ref{section_2} reviews the governing equations of the Extended Single Particle Model (ESPM). Section \ref{section_3} details the proposed Adaptive Ensemble Sparse Identification (AESI) method, covering ensemble learning, evolutionary optimization, and experimental setup. Section \ref{section_4} explains the conformal prediction approach for uncertainty quantification of AESI predictions. Section \ref{section_5} presents results on model performance, feature analysis, and uncertainty assessment. Section \ref{section_6} concludes the study and suggests future research directions.

\section{Overview of LiB Cell Model Equations} \label{section_2}

The ESPM is a control-oriented model that captures the essential electrochemical behavior of li-ion batteries. It improves upon the traditional Single Particle Model (SPM) by incorporating electrolyte dynamics, which are crucial for accurate voltage prediction \cite{fan2016modeling}. The ESPM simplifies the detailed Pseudo Two-Dimensional (P2D) model by representing each electrode as a single spherical particle, retaining the key physics of lithium intercalation and deintercalation within the solid phase. 

The solid-phase lithium concentration $ c_{s,i}$ (where $i = n,p$) dynamics is governed by Fick’s second law:

\begin{equation} 
\frac{\partial c_{s,i}(r, t)}{\partial t} = D_{s,i} \frac{\partial^2 c_{s,i}(r, t)}{\partial r^2} + \frac{2 D_{s,i}}{r} \frac{\partial c_{s,i}(r, t)}{\partial r}
\label{eq:ficks_law}
\end{equation}

\noindent where $r$ is the radial coordinate within the particle. %The associated boundary conditions are:
%\begin{equation} \label{eq:130}
%\left. \frac{\partial c_{s,i}(r, t)}{\partial r} \right|_{r=0} = 0 \quad
%\end{equation}
%\begin{equation} \label{eq::140}
%\quad D_{s,i} \left. \frac{\partial c_{s,i}(r, t)}{\partial r} \right|_{r=R_i} = -\frac{J_i(t)}{F}
%\end{equation}

The dynamics of lithium concentration in the electrolyte phase is described by:
\begin{equation}
\epsilon_{e,i} \frac{\partial c_{e}(x,t)}{\partial t} = D_{e,i} \frac{\partial^2 c_{e}(x,t)}{\partial x^2} + \frac{a_i (1 - t_0^+)}{F} J_i(t)
\label{eq:ficks_law_liquid}
\end{equation}
\noindent where $\epsilon_{e,i}$ is the porosity, $D_{e,i}$ the effective diffusion coefficient, $c_e$ the electrolyte concentration, $t_0^+$ the transference number, $x$ the spatial coordinate, and $a_i$ is the specific surface area. For brevity, the boundary conditions associated with Equations \eqref{eq:ficks_law} and \eqref{eq:ficks_law_liquid} are provided in \cite{fan2016modeling}.%The boundary conditions for the electrolyte concentration are:
% \begin{equation}
% \left. \frac{\partial c_e(x,t)}{\partial x} \right|_{x=0} = 0, \quad \left. \frac{\partial c_e(x,t)}{\partial x} \right|_{x=L} = 0
% \end{equation}

% % \begin{equation}
% % \left. \frac{\partial c_e(x,t)}{\partial x} \right|_{x=L} = 0
% % \end{equation}

% \begin{equation}
% c_e(x,t)\big|_{x=L_p^-} = c_e(x,t)\big|_{x=L_p^+}
% \end{equation}

% \begin{equation}
% c_e(x,t)\big|_{x=L_p + L_s^-} = c_e(x,t)\big|_{x=L_p + L_s^+}
% \end{equation}

% \begin{equation}
% -D_{e,p} \left. \frac{\partial c_e(x,t)}{\partial x} \right|_{x=L_p^-} = 
% -D_{e,s} \left. \frac{\partial c_e(x,t)}{\partial x} \right|_{x=L_p^+}
% \end{equation}

% \begin{equation}
% -D_{e,s} \left. \frac{\partial c_e(x,t)}{\partial x} \right|_{x=L_p + L_s^-} =
% -D_{e,n} \left. \frac{\partial c_e(x,t)}{\partial x} \right|_{x=L_p + L_s^+}
% \end{equation}

The kinetic overpotential $\eta_i$ is given by:
\begin{equation}
\eta_i(t) = \frac{{R}T}{\alpha F} \sinh^{-1} \left( \frac{J_i(t)}{2 i_{0,i}(t)} \right)
\end{equation}
\noindent where the exchange current density $i_{0,i}$ depends on the kinetic rate constant $K_k$, maximum electrode concentration $c_{s,\text{max},i}$, and average electrolyte concentration $c_e$:
\begin{equation}
i_{0,k}(t) = F K_k \sqrt{c_{s,\text{surf},k} \left( c_{s,\text{max},k} - c_{s,\text{surf},k} \right) c_{e}}
\end{equation}
\par The liquid phase potential $\phi_e$ \eqref{eq::180} is defined as a function of the effective ionic conductivity $\kappa^{eff}$, the effective diffusional conductivity  $\kappa_D^{eff}$, and the ionic current $i_e$ in the liquid phase.

\begin{equation} \label{eq::180}
\frac{\partial \phi_e (x, t)}{\partial x} = -\frac{i_e (t)}{\kappa^{eff}} + \frac{\kappa_D^{eff}}{\kappa^{eff}} \frac{\partial \ln c_e (x, t)}{\partial x}
\end{equation}
\par Last, the terminal voltage is expressed as:
\begin{equation} \label{eq:1}
\begin{split}
V_{espm}(t) = U_p(c_{s,p},t) - U_n(c_{s,n},t) - (\eta_p(0,t) \\ - \eta_n(L,t) ) - (\phi_{e}(0,t) - \phi_{e}(L,t)) - I(t)R_c
 \end{split}
\end{equation}

\noindent where $U_i$ denotes the open-circuit potential, and $R_0$ the ohmic resistance. In this work, the PDEs governing the solid-phase and electrolyte concentrations are approximated using the Galerkin projection method, as described in \cite{8008775}.

\begin{figure*}[t!] %\label{fig:1}
    \begin{center}
       \includegraphics[angle=0,scale=0.35]{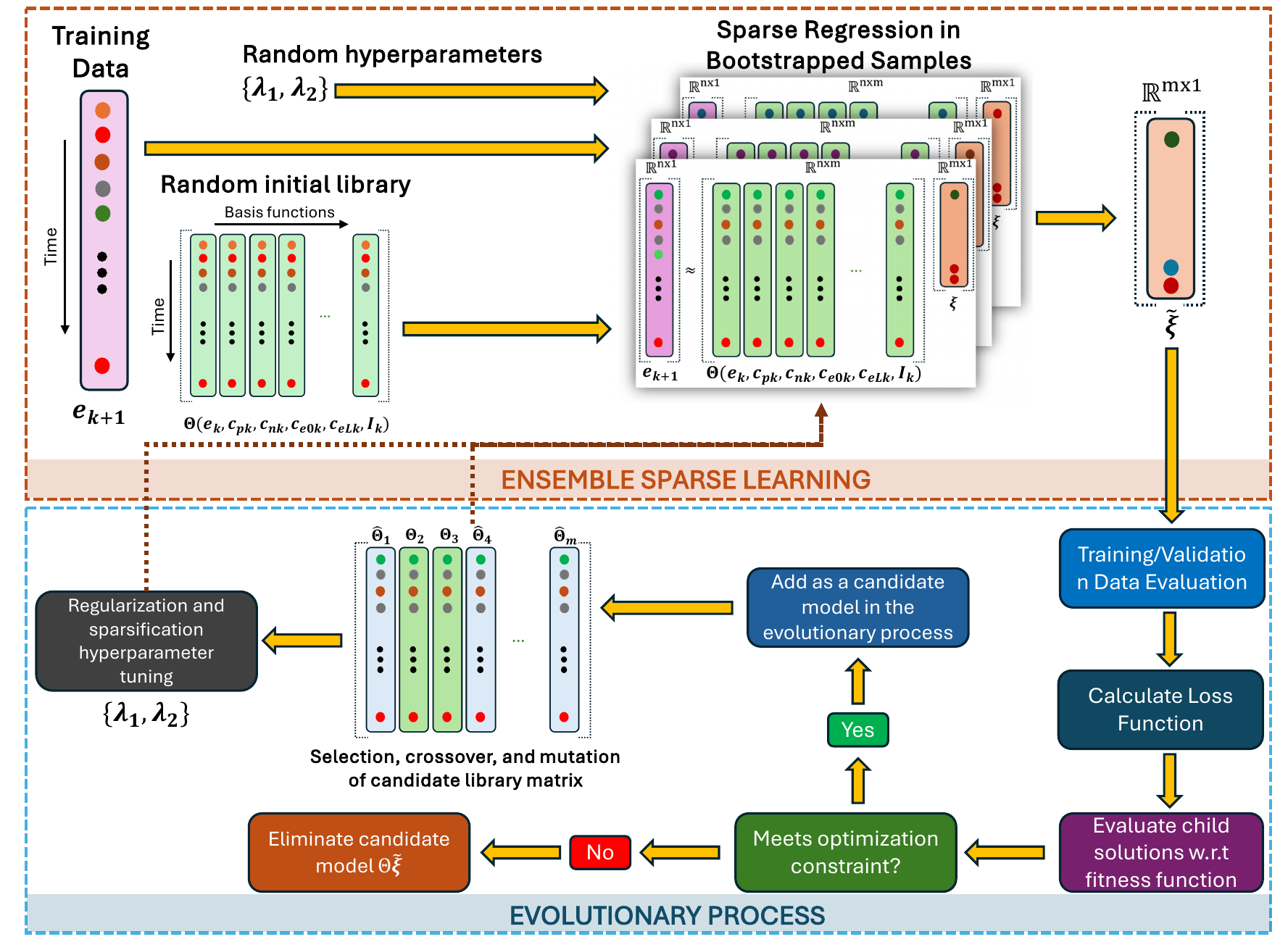}
    \caption{Schematic of the proposed Adaptive Ensemble Sparse Identification (AESI) framework.}
    \label{fig:schematic}
    \end{center}
\end{figure*}
\section{Adaptive Ensemble Sparse Identification Procedure} \label{section_3}
\begin{figure}[H] 
    \begin{center}
       \includegraphics[angle=0,scale=0.335]{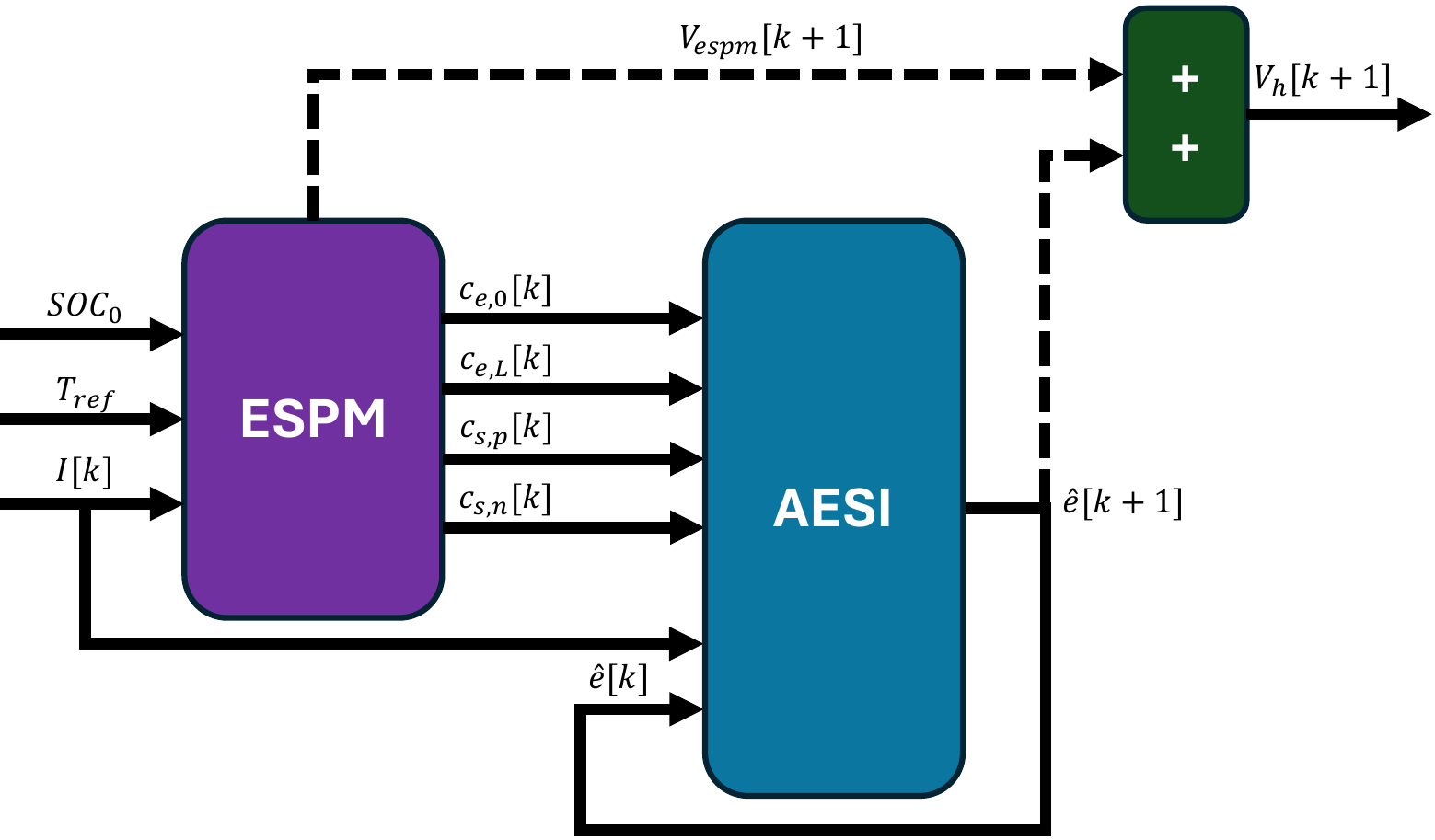}
    \caption{Hybrid li-ion battery model schematics: integration of ESPM with AESI model.}
    \label{fig:1_2}
    \end{center}
\end{figure}

The proposed (AESI) framework aims to augment the ROMs for LiBs by capturing missing dynamics. The method combines active sparse learning with evolutionary optimization to provide robust prediction model candidates. Figure \ref{fig:schematic} shows the schematic of the proposed method. To enhance the predictive accuracy of the ROM, a hybrid architecture, as illustrated in Fig. \ref{fig:1_2}, is introduced to augment the ESPM with a data-driven component trained to capture residual dynamics not modeled by the physics-based equations. Specifically, the hybrid model $V_h$ predicts the terminal voltage at time step $k+1$ as:
\begin{equation}
\begin{split}
V_h[k+1] = V_{\text{ESPM}}[k+1] + f_{\text{AESI}}(e[k], I[k], c_{s,p}[k], \dots \\ c_{s,n}[k], c_{e,0}[k], c_{e,L}[k])
\label{eq:hybrid_voltage}
\end{split}
\end{equation}
where $f_{\text{AESI}}(\cdot)$ is the proposed ML model that infers unmodeled dynamics from historical inputs and latent physical states. This missing dynamics is reflected in the voltage error $e$, quantifying the discrepancy between experimental measurements $V_{exp}$ and the physics-based prediction $V_{espm}$. The error is decomposed as $e(t) = \hat{e}(t) + \epsilon(t)$, where $\hat{e}_{k+1} = f_{\text{AESI}}(e_k, I_k,\dots)$ is the predicted error by the AESI model and $\epsilon$ is a prediction residual. The model is formulated based on the SINDy approach \cite{brunton2016discovering} to identify a representation of the error dynamics. The discrete-time formulation is given by:
% The main goal of the model is to learn the voltage error dynamics expressed as follows:
% \begin{equation} 
% \label{eq:20}
% e(t) = V_{exp}(t) - V_{espm}(t)
% \end{equation}
% \noindent where $V_{exp}$ denotes the experimental voltage and $V_{espm}$ the predicted voltage from the PBM at time $t$. The error is decomposed as $e(t) = \hat{e}(t) + \epsilon(t)$, where $\hat{e}$ is the reconstructed error predicted by the ML model and $\epsilon$ is a residual. The model is formulated based on the SINDy approach \cite{brunton2016discovering}, which captures the dynamical functional relationship between the error at the next time step and past error values, as well as input variables. The discretized system is expressed as:
\begin{equation} \label{eq:error_sindy}
\begin{split}
   \hat{e}[k+1] = \Theta(e[k],I[k],c_{s,p}[k],c_{s,n}[k],c_{e,0}[k], c_{e,L}[k])\xi 
\end{split}
\end{equation}
where $\Theta(\cdot)$ corresponds to a library operator and $\xi \in \mathbb{R}^{m}$ is a sparse coefficient vector. The matrix $\Theta \in \mathbb{R}^{n \times m}$ is constructed from a diverse set of potential basis functions, as outlined in Eq. \ref{eq:theta_split}. From this large dictionary, a subset of candidate predictors $\Theta_s \subseteq \Theta$ (Eq. \ref{eq:lib_set}) is then selected using an evolutionary algorithm. This feature pre-selection step reduces the dimensionality of the sparse regression problem, facilitating a more efficient search in the function space, as least squares regression scales asymptotically based on the library size %($\mathcal{O}(m^3)$ where $m$ is the number of functions, i.e. columns, in $\Theta$)
\cite{quade2018sparse}. This approach contrasts with conventional SINDy implementations that operate on a fixed library, thereby improving flexibility and ultimately accuracy in predicting complex nonlinear behaviors.

\vspace{-1.5em} 
\begin{equation} \label{eq:theta_split}
\begin{split}
\mathbf{\Theta}(\mathbf{e}, \mathbf{U}) = \Big[\, \mathbf{1},~ 
&\underbrace{T_1(\mathbf{e}),\, T_2(\mathbf{e}),\, \ldots,\, T_n(\mathbf{e})}_{\text{State Chebyshev polynomial terms}}, \\
&\underbrace{T_1(\mathbf{U}_1),\, \ldots,\, T_n(\mathbf{U}_1),\, \ldots,\, T_n(\mathbf{U}_m)}_{\text{Input Chebyshev polynomial terms}},\, \ldots, \\
&\underbrace{T_i(\mathbf{e}) T_j(\mathbf{U}_1),\ T_i(\mathbf{e}) T_j(\mathbf{U}_2) T_k(\mathbf{U}_3)}_{\text{Cross Chebyshev polynomial terms}}, \, \ldots,\, \\
& \underbrace{\phi(\mathbf{e}, \mathbf{U})}_{\text{Additional terms}} 
\,\Big]
\end{split}
\end{equation}
\noindent \vspace{-1.0em} 
\begin{equation} \label{eq:lib_set}
\mathbf{\Theta_s} = \left[ \begin{array}{c c c c c}
\vert & \vert &  & \vert & \vert \\
\mathbf{\Theta_{s,1}} & \mathbf{\Theta_{s,2}} & \cdots & \mathbf{\Theta_{s,m-1}} & \mathbf{\Theta_{s,m}} \\
\vert & \vert &  & \vert & \vert
\end{array} \right]
\end{equation}

In the library matrix, $T_n(\cdot)$ represents a Chebyshev polynomial function of the first kind of order $n$ (Eq. \ref{chebyshev_func}), $\mathbf{U_k}$ is the k$^{th}$ input variable from the set $\mathbf{U} = \{ I, c_{s,p}, c_{s,n}, c_{e,0}, c_{e,L}\}$, and product of Chebyshev polynomials from state/input variables is constrained by $i + j + k \leq d$, for the maximum polynomial degree $d$ defined by the selective process of library terms. The adoption of Chebyshev polynomials is motivated by their numerical stability and orthogonality, which enhance approximation performance and reduce multicollinearity among the basis functions \cite{mostajeran2025scaled}.  Additionally, $\phi(\mathbf{e}, \mathbf{U})$ represents possible additional nonlinear basis functions. In this formulation, trigonometric and hyperbolic functions are selected as potential terms for $\phi(\mathbf{e}, \mathbf{U})$ based on their ability to capture more complex nonlinear processes that govern the error dynamics \cite{AHMADZADEH2024110743,da2024improving}. 
\begin{equation} \label{chebyshev_func}
\begin{aligned}
T_1(x) &= x \\
T_2(x) &= 2x^2 - 1 \\
&\vdots \\
T_{n+1}(x) &= 2xT_n(x) - T_{n-1}(x)
\end{aligned}
\end{equation}

As shown in Eq. \ref{eq20}, all features are scaled to the range $[-1,1]$ to promote numerical stability,  uniform regularization, and improved model interpretability. 
\begin{equation} \label{eq20}
U' = 2 \cdot \frac{U - U_{\min}}{U_{\max} - U_{\min}} - 1,~~ e' = 2 \cdot \frac{e - e_{\min}}{e_{\max} - e_{\min}} - 1
\end{equation}

The identification of $\xi = (\xi_1, \xi_2,\dots,\xi_m)^T$ can be formulated as a regularized optimization problem solved by sequentially thresholded ridge (STRidge) regression \cite{rudy2019data}:
\begin{equation}  \label{eq:SINDyC}
\begin{split}
    \hat{\xi}^{\lambda_1,\lambda_2} = \arg \min_{\xi \in \mathbb{R}^{m}} \left\| \mathbf{{e}}[k+1] - \Theta(\mathbf{{e}}[k],\mathbf{U}[k])\xi \right\|_{2} + \lambda_{1} \left\| \xi \right\|_{2} 
\end{split}
\end{equation}

\noindent which is followed by a thresholding rule:
\begin{equation}  \label{eq:SINDyC1}
\begin{split}
\hat{\xi}_j^{(l+1)} = 
\begin{cases}
\hat{\xi}_j^{(l)}, & \text{if } |\hat{\xi}_j^{(l)}| \geq \lambda_2 \\
0, & \text{otherwise}
\end{cases}
\quad \forall ~ j = 1, \dots, m
\end{split}
\end{equation}
\noindent where $\lambda_1$ represents the L2-norm regularization parameter that provides uniform coefficient shrinkage, addressing multicollinearity and model overfitting issues. The elimination of unnecessary basis functions is controlled by the sparsification hyperparameter $\lambda_2$, whose range (defined as  [$10^{-2}$, $5$]) is based on the minimum value of $\lambda_2$ that will set the model coefficients to zero when the library matrix contains all possible combinations of basis functions. The regularization parameter $\lambda_1$ varies in the range [$10^{-13}$, $10^{-1}$]. Additionally, the individual polynomial function maximum order $p_{c,i}$ ranges from [1, 5], constrained by the maximum total polynomial degree $d$, which is set within the range [1, 5] and satisfies $p_{c,1} + p_{c,2} + \dots + p_{c,6} \leq d$.

% As illustrated by Fig. \ref{fig:1_2}, the hybrid battery model output $V_{h}$ can be expressed as follows:
% \begin{equation} \label{eq:20}
% V_{h}(t) = V_{espm}(t) + \hat{e}(t)
% \end{equation}
% where $\hat{e}$ is the predicted error by the proposed AESI framework.

% To enhance the predictive accuracy of the reduced-order model, a hybrid architecture, as illustrated in Fig. \ref{fig:1_2}, is then introduced to augment the ESPM with a data-driven component trained to capture residual dynamics not modeled by the physics-based equations. Specifically, the hybrid model predicts the terminal voltage at time step $k$ as:
% \begin{equation}
% \begin{split}
% V_h[k+1] = V_{\text{ESPM}}[k+1] + f_{\text{AESI}}(e[k], I[k], c_{s,p}[k], \dots \\ c_{s,n}[k], c_{e,0}[k], c_{e,L}[k])
% \label{eq:hybrid_voltage}
% \end{split}
% \end{equation}
% where $f_{\text{AESI}}(\cdot) = \Theta(\cdot)\xi$ is the proposed ML model that infers unmodeled dynamics from historical inputs and latent physical states.

% \begin{table}[h]
% \centering
% \caption{Hyperparameter range}
% \label{table:1}
% \begin{tabular}{lcc}
% \hline
% \textbf{Parameter} & \textbf{Range} & \textbf{Type} \\ 
% \hline
% $\lambda_1$; $\lambda_2$             & $[10^{-13},10^{-1}]$; $[10^{-2},5]$ & Continuous\\
% $p_c$;   $d$          & $[0,5]$; $[1,5]$  & Integer \\
% \hline
% \end{tabular}
% \end{table}
\subsection{Ensemble sparse learning}
The proposed framework adopts Moving Block Bootstrap (MBB) to improve the model robustness. For each bootstrap iteration, the training data set is resampled in contiguous blocks, preserving short-term temporal dependencies within the data. This resampling strategy is important for dynamical systems with temporal dependencies, as stationary bootstrap, based on the assumption of i.i.d. behavior, may potentially lead to biased model estimates \cite{lahiri2013resampling}. 

\begin{figure}[h!] 
    \begin{center}
       \includegraphics[angle=0,scale=0.5]{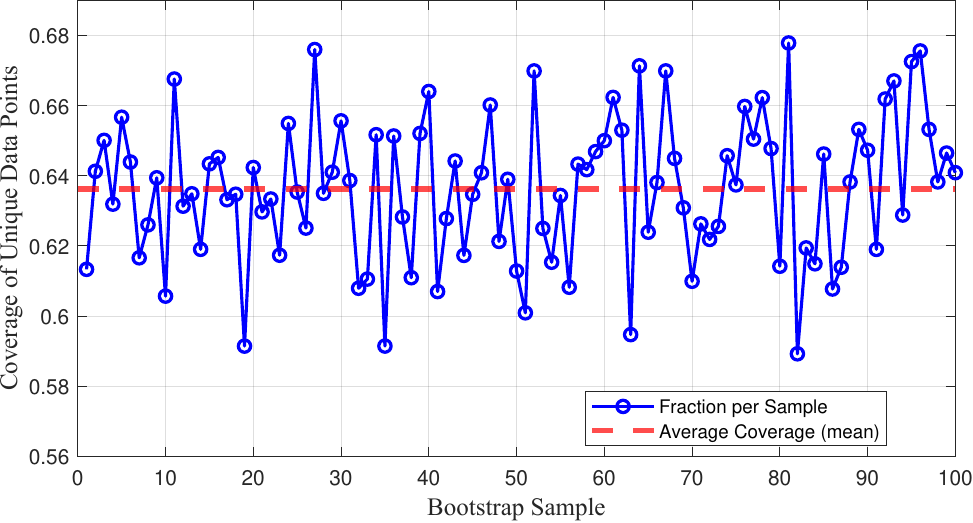}
    \caption{Coverage of unique points per MBB sample.}
    \label{fig:2}
    \end{center}
\end{figure}

The expected number of included training data points per MBB sample $\mathbb{E}[S_b]$ can be expressed as:
\begin{equation}
\mathbb{E}[S_b] \approx \sum_{i=1}^{n} \left[ 1 - \left(1 - \frac{1}{n} \right)^{\frac{n}{B}} \right] = n\left[ 1 - \left(1 - \frac{1}{n} \right)^{\frac{n}{B}} \right]
\end{equation}
\noindent where $n$ is the number of data points from the original training data set, $B$ the block size. For the standard bootstrap with replacement, as $n$ approaches to infinity, the expected number of unique points per sample can be approximated to 63.2\%, that is $\lim_{n \to \infty} \left[ 1 - \left(1 - \frac{1}{n} \right)^{n} \right] \approx 0.632$ \cite{lahiri2013resampling,fasel2022ensemble}. For a large $n$ and $n >> B$, $\mathbb{E}[S_b]$ tends to converge to a comparable coverage ratio. Figure \ref{fig:2} illustrates a random selection of bootstrapped samples for $B = 500$ and $n = 88,200$ (parameter values used in the current study), where the average coverage ratio is reported as approximately 63.6\%. 

\subsubsection{Top model selection} 

As part of the bootstrap-based ensemble strategy, 100 resampled datasets are generated from the original training set using the Moving Block Bootstrap (MBB). For each resampled dataset, the STRidge regression problem defined in Eqs. \ref{eq:SINDyC}-\ref{eq:SINDyC1} is solved, yielding a set of sparse coefficient vectors $\xi^{(b)}$, where $b$ indexes the b$^{th}$ bootstrapped sample. Among these generated models, the top 10\%  \cite{fasel2022ensemble} are selected based on their out-of-bag (OOB) prediction performance \cite{hastie2009elements}. The OOB mean squared error (MSE) for the b$^{th}$ model is computed as follows:
\begin{equation}
\text{MSE}^{(b)}_{\text{OOB}} = \frac{1}{|\mathcal{C}^{-i}|} \sum_{j \in \mathcal{C}^{-i}} \left(e_j - \Theta^{(b)}(\hat{e}_j, U_j) \hat{\xi}^{(b)} \right)^2
\end{equation}
\noindent where $\mathcal{C}^{-i}$ denotes the out-of-bag dataset for the b$^{th}$ bootstrap sample $\mathbb{Z}_{b}$, that is:
\begin{equation}
\mathcal{C}^{-i} = \left\{ b_j : (e_j, U_j) \in \mathbb{Z}_{b}^{-} \right\}
\end{equation}

After selecting the top-performing models, the final sparse representation is obtained using two distinct strategies, namely bootstrap aggregation (bagging) and stability selection based on inclusion probability.

\subsubsection{Bagging} In this approach, the final sparse vector $\hat{\xi}^{bag}$ is computed by averaging the coefficients from the $M$ selected top-performing models (as defined in Eq. \ref{bagging}), where each $\xi^{(b)}$ is trained on a different bootstrapped model. This approach tends to reduce model variance and improve robustness by preventing overfitting to individual samples \cite{hastie2009elements}.
\begin{equation} \label{bagging}
\hat{\xi}^{bag} = \frac{1}{M} \sum_{b=1}^{M} \xi^{(b)}
\end{equation}
\subsubsection{Stability selection} This structure-learning technique evaluates the relevance of predictors by quantifying how consistently they are selected across multiple resampled models \cite{meinshausen2010stability}. For the j$^{th}$ element of the sparse coefficient vector $\hat{\xi}^{(b)}$ (i.e., j$^{th}$ basis function in the library function $\Theta$), the inclusion probability $\hat{\pi}_j$ is defined as follows:
% Stability selection is a structure-learning procedure that evaluates the relevance of library functions by quantifying how consistently they are selected across multiple resampled models \cite{meinshausen2010stability}. The inclusion probability $\hat{\pi}_j$ for the j$^{th}$ element of the sparse vector $\hat{\xi}^{(m)}$ (that corresponds to the j$^{th}$ basis function of $\Theta$ matrix) can be calculated as follows:
\begin{equation}
\hat{\pi}_j = \mathbb{P}\left(\hat{\xi}_j^{(b)} \neq 0\right) \approx \frac{1}{M} \sum_{b=1}^{M} \mathds{1}\left( \hat{\xi}_j^{(b)} \neq 0 \right)
\end{equation}
\noindent where $M$ is the number of selected bootstrap models and $\mathds{1}(\cdot)$ is the indicator function. A term is retained in the final model if its estimated inclusion probability exceeds a threshold $\tau$, otherwise it is discarded. This selection procedure can be represented as:

\begin{equation}
\Theta_{\text{j}}^{sel} = \left[\, \Theta_j \;\middle|\; \hat{\pi}_j > \tau \,\right]
\label{eq:theta_sel}
\end{equation}

\noindent where $\Theta_j$ denotes the j$^{th}$ column of the library matrix, and $\Theta^{sel}$ contains only the columns associated with stable terms. The regularized optimization is then performed using the reduced library to compute the final vector $\hat{\xi}^{\text{sel}}$ \cite{fasel2022ensemble}:

\begin{equation}
\begin{split}
    \hat{\xi}^{\text{sel}} = \arg\min_{\xi \in \mathbb{R}^{s}} \left\| e[k+1] - \Theta^{sel}(\hat{e}[k], U[k]) \xi^{\text{sel}} \right\|_{2} + \lambda_1 \|\xi^{\text{sel}}\|_{2} 
\end{split}
\end{equation}

\noindent followed by a thresholding rule:
\begin{equation}  \label{eq:21}
\begin{split}
\hat{\xi}_j^{\text{sel}(l+1)} = 
\begin{cases}
\hat{\xi}_j^{\text{sel}(l)}, & \text{if } |\hat{\xi}_j^{\text{sel}(l)}| \geq \lambda_2 \\
0, & \text{otherwise}
\end{cases}
\quad \forall ~ j = 1, \dots, m
\end{split}
\end{equation}
\noindent where $\hat{\xi}^{\text{sel}}_j$ denotes the refitted coefficient estimated over the original training dataset using only the most stable basis functions. The threshold $\tau$ is adaptively selected through the evolutionary algorithm to balance sparsity and predictive performance. 

\subsubsection{Final ensemble model} The final model is then obtained by aggregating the information from the selected ensemble of sparse models, identified through the techniques described above. The modified voltage error predictor is expressed as:
\begin{equation} \label{eq:error_sindy_final}
\begin{split}
   \hat{e}[k+1] \approx \Theta(e[k],U[k])\xi^{ens} 
\end{split}
\end{equation}
\noindent where $\xi^{ens}$ represents final sparse coefficient vector derived via ensemble learning ($\xi^{ens} = \xi^{bag}$ for bagging or $\xi^{ens} = \xi^{sel}$ for stability selection).

\subsection{Evolutionary-based selection}
The prediction model is selected based on a hybrid loss function $\mathcal{L}(\xi,\Theta_s)$. The first component, the out-of-bag (OOB) loss $\mathcal{L}^{\text{OOB}}_{\text{boot}}$, quantifies the average prediction error over the OOB samples from the top-selected bootstrap models, as defined in Eq. \ref{loss_OOB}. To enhance generalization and account for model performance under unseen conditions (i.e., outside the boostrapped training distribution), a validation loss term $\mathcal{L}_{\text{valid}}$ is incorporated (Eq. \ref{loss_valid}). 
\begin{equation} \label{loss_OOB}
\mathcal{L}^{\text{OOB}}_{\text{boot}} = \frac{1}{N_B} \sum_{b=1}^{N_B} \text{MSE}^{(b)}_{\text{OOB}}
\end{equation}
\begin{equation} \label{loss_valid}
\mathcal{L}_{\text{valid}} = \frac{1}{|\mathcal{D}_{val}|} \sum_{j \in \mathcal{D}_{val}} \left(e_j - \Theta(\hat{e}_j, U_j) \hat{\xi} \right)^2
\end{equation}
$\mathcal{D}_{val}$ represents the index set of the validation dataset. To promote model sparsity and prevent overfitting, a third term $\mathcal{L}_{\text{sparse}}$ penalizes the number of active basis functions in the coefficient vector $\xi$. 
\begin{equation}
\mathcal{L}_{\text{sparse}} = \|\xi\|_{0} 
\end{equation}

The total loss $\mathcal{L}$ can be then defined as follows:
\begin{equation}
\mathcal{L}(\xi,\Theta_s) = \gamma_1\mathcal{L}^{\text{OOB}}_{\text{boot}} + \gamma_2\mathcal{L}_{\text{valid}} + \gamma_3 \mathcal{L}_{\text{sparse}}
\end{equation}

In the genetic algorithm, each candidate solution is associated with a genome \( P_i \) that encodes a selected library matrix \( \Theta_s \), along with regularization \( \lambda_1 \) and sparsification \( \lambda_2 \) parameters.
\begin{equation}
P_i = \{ \Theta_{s}, \lambda_1, \lambda_2\}
\end{equation}

The fitness score is maximized according to the optimization problem described in Eq. \ref{eq:fit},
\begin{equation} \label{eq:fit}
\begin{split}
   \{ \lambda_1, \lambda_2, \Theta_s \} = \arg \max_{\lambda_1, \lambda_2, \Theta_s} \left( 1 - \mathcal{L}(\xi, \Theta_s) \right)  \\ \quad \text{s.t.:} \quad \rho(e, \hat{e})_{train} > \epsilon_1, \quad\rho(e, \hat{e})_{valid} > \epsilon_2  \\ \lambda_1 \in \{10^{-13}, 10^{-1}\},  
    \lambda_2 \in \{10^{-2}, 5\} \\ 
   p_c \in \{0, 5\},
    d \in \{1, 5\},
    \tau \in \{0.3, 0.7\} 
\end{split}
\end{equation}
\noindent where $\epsilon_1$ and $\epsilon_2$ represent a minimum Pearson correlation for the training and validation sets, respectively.

%% losses here!

% \textcolor{red}{STRidge regression: Formulation of the problem}

% \textcolor{red}{Ensemble learning - resampling technique}

% \textcolor{red}{Evolutionary process to select basis functions}
% ...

\subsection{Data collection}

\begin{figure}[h!]
\centering
\subfigure[Current profile]{
    \includegraphics[width=0.475\textwidth]{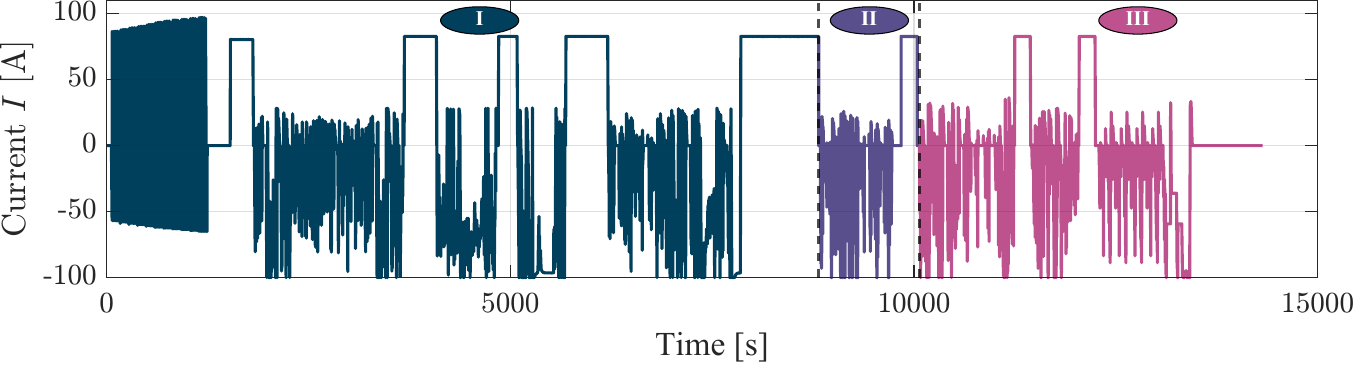}
}
\subfigure[Voltage profile]{
    \includegraphics[width=0.475\textwidth]{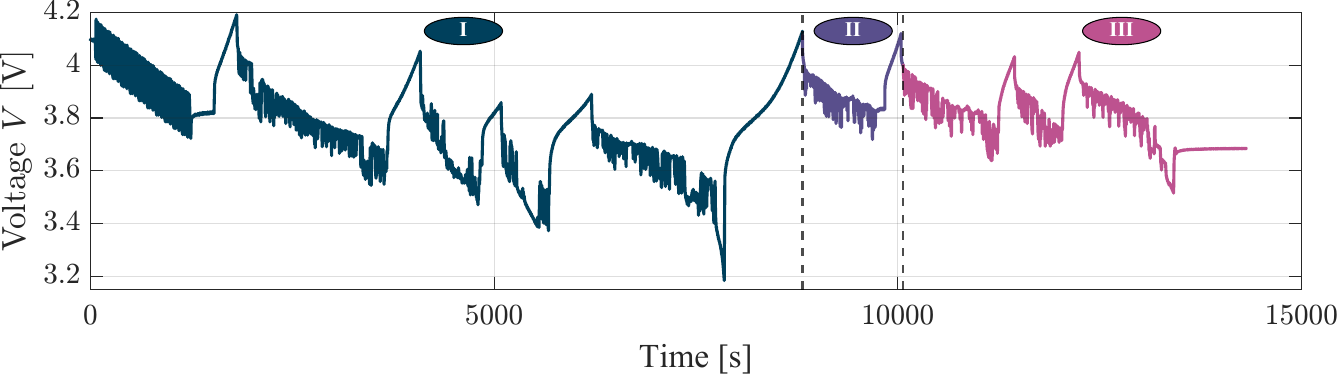}
}
\caption{Current and voltage profiles for training (I), validation set (II), testing (III) sets.}
\label{fig:profiles}
\end{figure}

\begin{figure*}[!h]
\centering
\subfigure[Training set]{
    \includegraphics[width=0.3\textwidth]{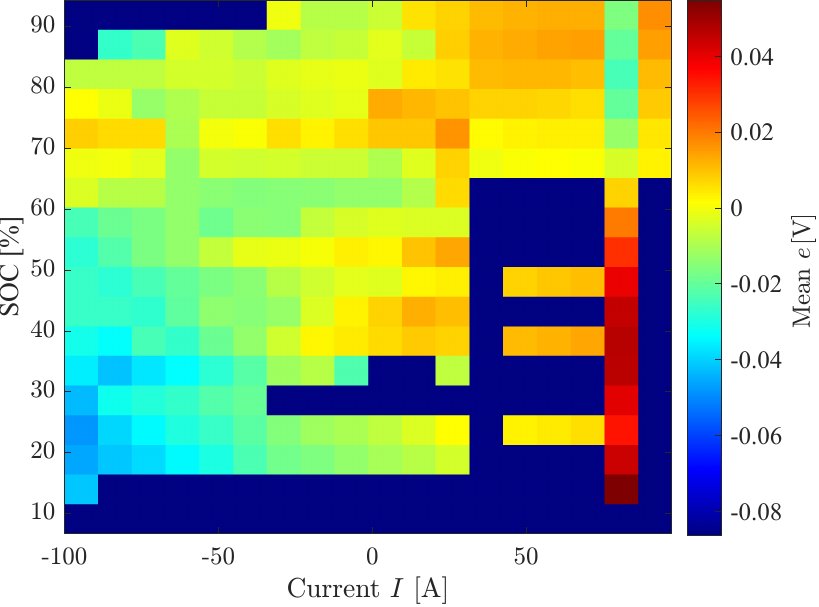}
}
\subfigure[Validation set]{
    \includegraphics[width=0.315\textwidth]{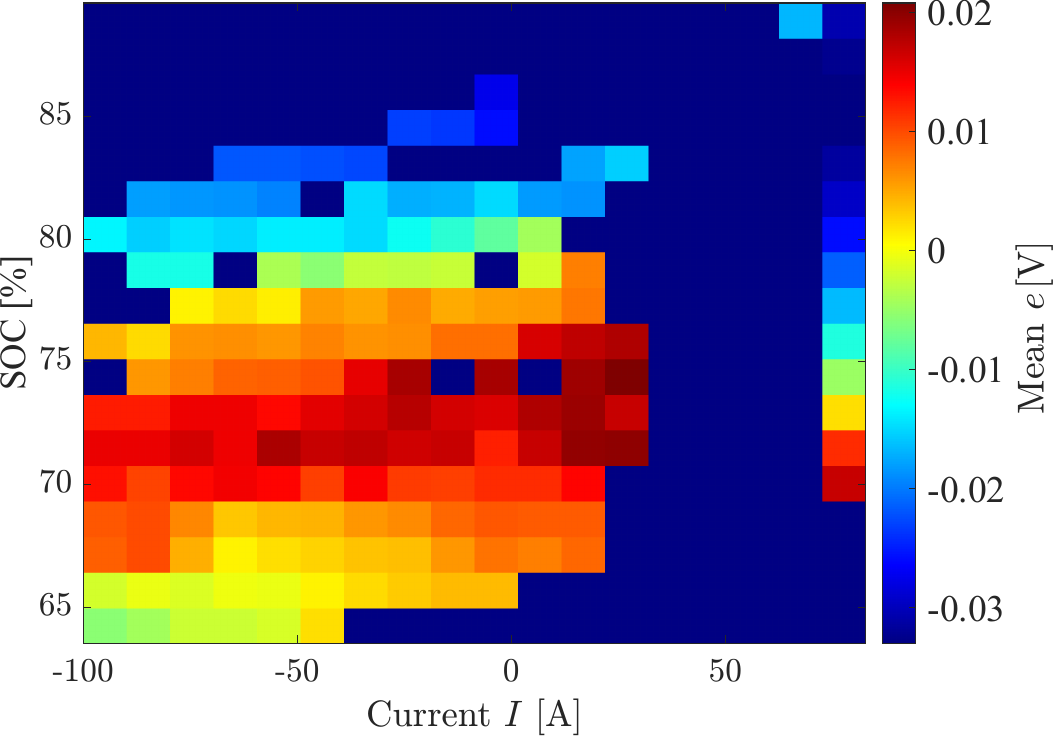}
}
\subfigure[Testing set]{
    \includegraphics[width=0.315\textwidth]{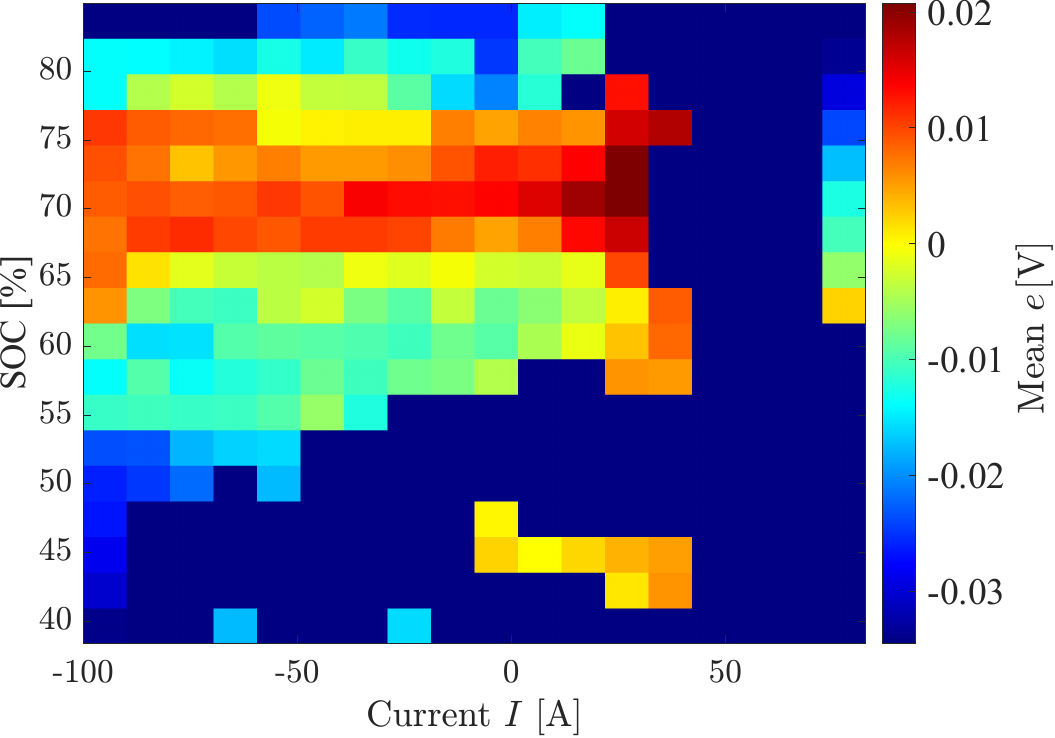}
}
\caption{Voltage error $e$ distribution of ESPM across $(I, \mathrm{SOC})$ bins.}
\label{fig:error_map}
\end{figure*}

The training, validation, and testing data are obtained from experimental measurements on a 30 Ah nickel–manganese–cobalt (NMC) graphite cell, sampled at \mbox{10 Hz}. The training current profile $I$ is derived from a simulated battery electric vehicle (BEV) model operating under different standard driving cycles, including UDDS, HWFET, US06, and WLTC. To prevent over-discharge, a 5C charging pulse is applied after each cycle. The model is validated using the Artemis Urban cycle, while SC03, Japan Cycle (JC), and NEDC cycles are used for testing. Figure~\ref{fig:profiles} presents the current and voltage trajectories associated with each section, and Figure \ref{fig:error_map} shows the mean voltage error $e$ distribution of ESPM across ($I$,$SOC$) bins.

\section{Uncertainty Quantification of AESI with Conformal Prediction} \label{section_4}
%For the proposed AESI framework, accurate prediction alone is not sufficient and valid confidence bounds around predictions are also required to ensure the reliability of the voltage error predictions. 
%Conformal prediction (CP) is integrated with AESI to provide a model-agnostic and distribution-free framework that constructs prediction intervals satisfying user-specified coverage levels. %In this section, a brief overview of the conformal prediction framework will be provided, followed by a description of Sequential Predictive Conformal Inference (SPCI) \cite{SPCI-Hu}, a state-of-the-art conformal prediction method tailored for non-exchangeable time series data.
%\subsection{Conformal Prediction Overview}
%Conformal prediction is an effective method to provide valid prediction intervals for the prediction models, including neural networks, without making any assumptions about the prediction model and underlying data distribution. Conformal prediction has been used in different real-world applications, including wind energy \textcolor{red}{[cite]} and vehicle trajectory \textcolor{red}{[cite]} due to its generality, simplicity, and theoretical guarantees. 

Conformal prediction (CP) is integrated with AESI to provide a model-agnostic and distribution-free framework that constructs prediction intervals satisfying user-specified coverage levels. CP constructs predictive intervals around point estimates using past residuals as \textit{non-conformity scores}, providing statistical guarantees without making strong assumptions about the data distribution or model form. For regression problems, a common choice of non-conformity score is the prediction residual, $\hat{\epsilon}_t = Y_t - \hat{Y}_t$, where $\hat{Y}_t$ is the predicted value of the output variable $Y_t$.

In the traditional \textit{split conformal} method \cite{splitconformal}, the data are assumed exchangeable, and a fixed calibration set is used to determine quantiles of residuals for constructing prediction intervals:
\begin{equation}
\hat{C}_{t-1}(X_t) = \left[\hat{Y}_t + q_{\alpha/2},\ \hat{Y}_t + q_{1-\alpha/2} \right],
\end{equation}
where $\alpha$ denotes the desired significance level or error rate, $q_{1-\alpha/2}$ denotes the $(1 - \alpha/2)$ quantile of the residuals on the calibration set and $t-1$ indicates the prediction interval is constructed using previous up to $t-1$ many observations. 

While classical CP guarantees marginal coverage $P(Y_t \in \hat{C}_{t-1}(X_t)) \geq 1 - \alpha$ under the data exchangeability assumption, it cannot guarantee conditional coverage $P(Y_t \in \hat{C}_{t-1}(X_t) | X_t)$, and its validity breaks down under temporal dependence, which is inherent to time series data \cite{SPCI-Hu}, \cite{EnbPI-Hu}.
To address the limitations in the presence of time series data, \textit{Sequential Predictive Conformal Inference} (SPCI) is proposed in \cite{SPCI-Hu}. SPCI handles non-exchangeable residuals by adaptively estimating the conditional quantiles of prediction residuals while accounting for temporal dependencies.

The key idea of SPCI is to leverage recent $w$ residuals as features for quantile regression to forecast the conditional distribution of future residuals. Instead of using empirical quantiles directly, SPCI estimates residual quantiles using a quantile regression model, typically trained on a sliding window of past residuals:
\begin{equation}
\label{eq:cp}
\hat{C}_{t-1}(X_t) = \left[\hat{Y}_t + \hat{Q}_t(\hat{\beta}),\ \hat{Y}_t + \hat{Q}_t(1 - \alpha + \hat{\beta})\right]
\end{equation}
where $\hat{Q}_t(\cdot)$ is a learned quantile function and $\hat{\beta}$ is a calibration parameter to minimize the interval width and set as $\alpha/2$  similarly in \cite{SPCI-Hu}.

SPCI uses quantile regression forests (QRF) \cite{qrf} to estimate conditional quantiles of the residuals and retains the model-agnostic nature of CP while enhancing adaptivity to temporal correlations in prediction residuals \cite{SPCI-Hu}. Theoretical analysis in \cite{SPCI-Hu} shows that SPCI achieves asymptotic conditional coverage even under temporal dependencies, provided mild regularity conditions are met:
\begin{equation}
\lim_{T \to \infty} P(Y_t \in \hat{C}_{t-1}(X_t) \mid X_t) = 1 - \alpha
\end{equation}
\par In practice, SPCI produces significantly narrower intervals than traditional conformal methods for time series such as \textit{EnbPI}, especially when residuals exhibit strong autocorrelation \cite{EnbPI-Hu}. It also generalizes \textit{split conformal} and \textit{EnbPI}; if residuals are assumed to be independent and $\hat{Q}_t$ is the empirical quantile function, SPCI reduces to \textit{EnbPI} and if the quantile predictor is trained on the residuals from the calibration set, and residuals are not updated during prediction, SPCI reduces to \textit{split conformal}. SPCI thus provides a principled and flexible framework for prediction interval construction in time series models.

SPCI is integrated with AESI to construct valid and adaptive prediction intervals for future voltage error estimates. QRF is applied to form these intervals by using the residuals obtained from AESI ensemble predictions on the cascaded training and validation data set, as illustrated earlier in Fig. \ref{fig:profiles}. The significance level $\alpha$ in the SPCI method is set to 0.1 and the sliding-window length in the QRF algorithm is set to $w=200$, corresponding to 20 s of historical data. During inference, for each incoming instance in the testing data, the prediction interval is generated from quantiles estimated over the $w$ most recent residuals. These intervals are then updated online as new observations arrive, ensuring that the prediction bands remain both calibrated and informative. A step-by-step description of how SPCI quantifies uncertainty in AESI is provided in Algorithm \ref{alg:spci}.
\begin{algorithm}[H]
\caption{Uncertainty Quantification of AESI with SPCI}
\label{alg:spci}
\textbf{Input:} Cascaded training and validation data $\{(X_t, Y_t)\}_{t=1}^T$, testing data $\{(X_t, Y_t)\}_{t=T+1}^N$, trained AESI algorithm $\Theta \xi^{ens}$, significance level $\alpha$, QRF algorithm $\mathcal{Q}$

\textbf{Output:} Prediction intervals $\hat{C}_{t-1}(X_t)$ for testing data

\begin{algorithmic}[1]
\State Obtain prediction residuals $\hat{\varepsilon}$ using $\Theta \xi^{ens}$ and $\{(X_t, Y_t)\}_{t=1}^T$
\For{$t > T$ in $\{(X_t, Y_t)\}_{t=T+1}^N$} 
    \State Use QRF to obtain $\hat{Q}_t \gets \mathcal{Q}({\{\hat\varepsilon_t\}_{t=t-w+1}^t})$
    \State Obtain prediction interval $\hat{C}_{t-1}(X_t)$ as in \eqref{eq:cp}
    \State Compute new residual $\hat{\varepsilon}_t$
    \State Update residuals $\hat{\varepsilon}$ by sliding index forward
\EndFor
\end{algorithmic}
\end{algorithm}

\section{Results and Discussion} \label{section_5}
The performance of the proposed hybrid models was evaluated using separate training, validation, and testing datasets. The comparison includes two distinct ensemble strategies: AESI with bagging aggregation (AESI I) and AESI with stability selection (AESI II), as well as a baseline hybrid model employing SINDy with control (SINDy-C) combined with evolutionary algorithm \cite{da2024improving}. Through the optimization process, the optimal hyperparameters identified were: regularization parameter $\lambda_1 = 3.57 \times 10^{-12}$, threshold parameter $\lambda_2 = 1.743$, maximum degree of individual polynomials $p_c = 1$, maximum polynomial degree $d = 2$, and stability selection threshold $\tau = 0.41$. Table \ref{table:MSER} presents the Mean Squared Error Reduction (MSER), as defined in Eq.~\ref{eq:MSER}, which quantifies the improvement in voltage prediction accuracy relative to the original physics-based model (PBM). Among the evaluated approaches, AESI I achieved the highest MSER of 45.96\%, demonstrating superior generalization performance. AESI II yielded a lower MSER of 30.41\%, reflecting its more conservative model selection via stability filtering. In contrast, the SINDy-C hybrid model—which relies solely on sparse regression with tuned hyperparameters—offered the least improvement, with an MSER of 25.14\%. 
\begin{equation} \label{eq:MSER}
\text{MSER} = \left(\frac{\text{MSE}(V_{exp},V_{espm}) - \text{MSE}(V_{exp},V_{h})}{\text{MSE}(V_{exp},V_{espm})}\right)100\%
\end{equation}
% \begin{table}[h!]
% \centering
% \caption{Optimal hyperparameter values}
% \label{table:optimal_param}
% \begin{tabular}{lcc}
% \hline
% \textbf{Parameter} & \textbf{Optimal value}  \\ 
% \hline
% $\lambda_1$             & $3.57\times(10^{-12})$  \\
% $\lambda_2$     &  $1.743$    \\
% $p_c$             & $1$  \\
% $d$           & $2$   \\
% $\tau$           & $0.41$   \\
% \hline
% \end{tabular}
% \end{table}
Although the stability-based hybrid model (AESI II) achieves the highest MSER on the combined training and validation sets (88.21\%), its generalization to unseen data is notably lower than that of AESI I. This suggests that AESI II prioritizes the most dominant patterns in the training data, as it refits the final model using only frequently selected (stable) terms. In contrast, the bagging-based approach (AESI I) averages coefficients across an ensemble of sparse models, producing a smoother approximation that generalizes more effectively. As a result, AESI I achieves a higher MSER of 45.96\% on the test set. However, this improved generalization comes with a trade-off: AESI I yields a lower MSER (79.69\%) on the combined training and validation sets, reflecting a more conservative learning strategy. These results are consistent with the theoretical advantage of bagging in reducing variance and enhancing model robustness. Overall, AESI I provides a favorable balance between robustness and generalization in hybrid battery modeling.

\begin{figure}[!h]
\centering
\subfigure[ESPM]{
    \includegraphics[width=0.35\textwidth]{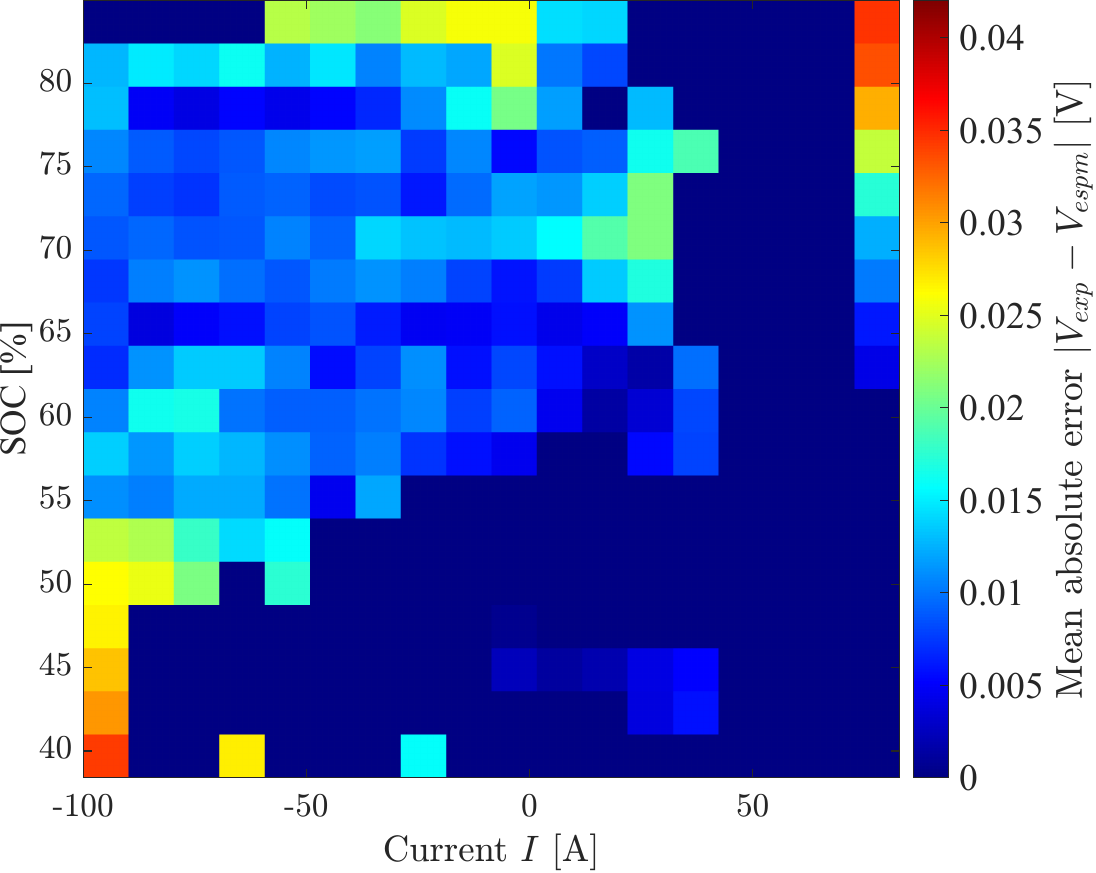}
}
\subfigure[ESPM + AESI I]{
    \includegraphics[width=0.35\textwidth]{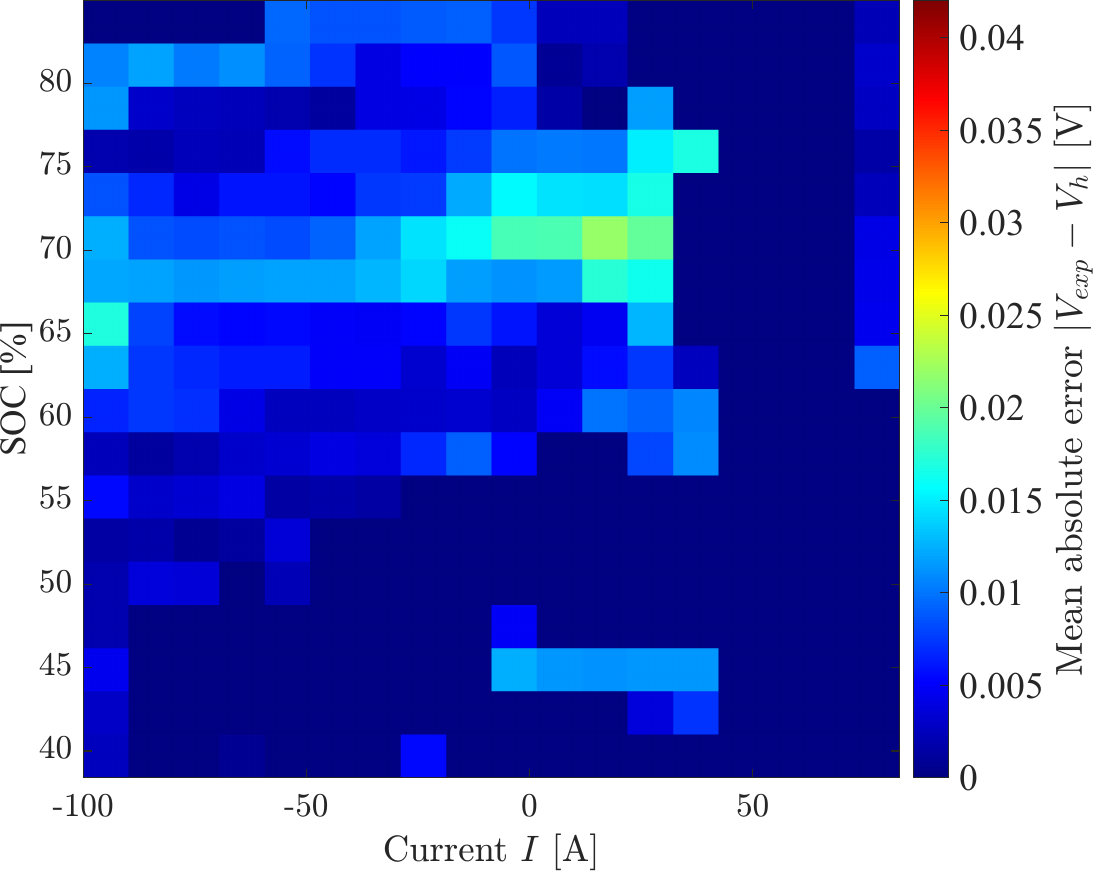}
}
\caption{Voltage absolute error distribution of ESPM and ESPM + AESI I hybrid model across $(I, \mathrm{SOC})$ bins.}
\label{fig:error_map_2}
\end{figure}

This improvement is further illustrated in Fig. \ref{fig:error_map_2}, which illustrates the absolute voltage error under testing conditions. The heatmap shows that AESI I effectively reduces large prediction errors, particularly in high C-rate regions, resulting in a more uniform error distribution across the operating domain.

% \begin{table}[h!]
% \centering
% \caption{Optimal hyperparameter values}
% \label{table:optimal_param}
% \begin{tabular}{ccccc}
% \hline
% \multicolumn{5}{c}{\textbf{Optimal value}} \\
% \hline
% $\lambda_1$ & $\lambda_2$ & $p_c$ & $d$ & $\tau$ \\
% $3.57\times 10^{-12}$ & $1.743$ & $1$ & $2$ & $0.41$ \\
% \hline
% \end{tabular}
% \end{table}

% \begin{figure}[h!]
% \centering
% \subfigure[Current profile]{
%     \includegraphics[width=0.44\textwidth]{plot_current.pdf}
% }
% \subfigure[Voltage profile]{
%     \includegraphics[width=0.44\textwidth]{plot_voltage.pdf}
% }
% \caption{Current and voltage profiles for training (I), validation set (II), testing (III) sets}
% \label{fig:profiles}
% \end{figure}

\begin{figure}[h!] 
    \begin{center}
       \includegraphics[angle=0,scale=.65]{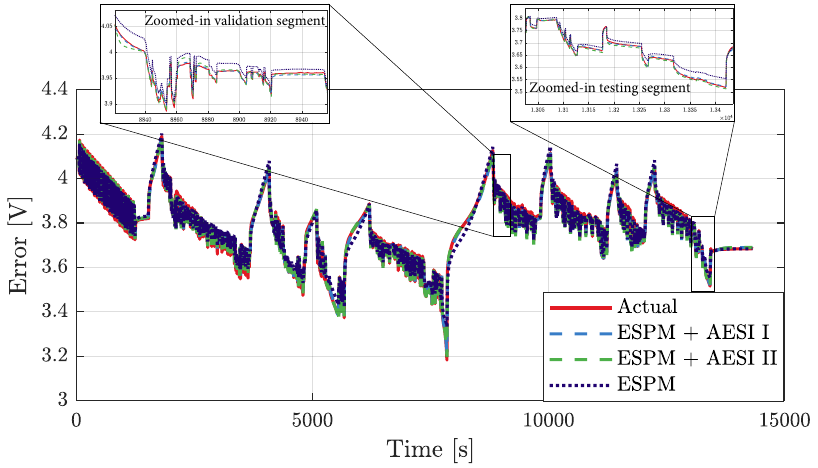}
    \caption{Voltage prediction of ESPM and hybrid model under training, validation and testing sets}
    \label{fig:plot_actual_hybrd}
    \end{center}
\end{figure}

\begin{table}[h!]
\centering
\caption{Comparison of voltage MSE reduction in testing set}
\label{table:MSER}
\begin{tabular}{lcc}
\hline
\textbf{Model} & \textbf{MSER [\%]} \\ 
\hline
Hybrid Model (ESPM + AESI I)         & $45.96 \%$  \\
Hybrid Model (ESPM + AESI II)      &  $30.41\%$   \\
Hybrid Model (ESPM + tuned SINDy-C)           & $25.14\%$  \\
\hline
\end{tabular}
\end{table}

\begin{figure}[h!] 
    \begin{center}
       \includegraphics[angle=0,scale=0.445]{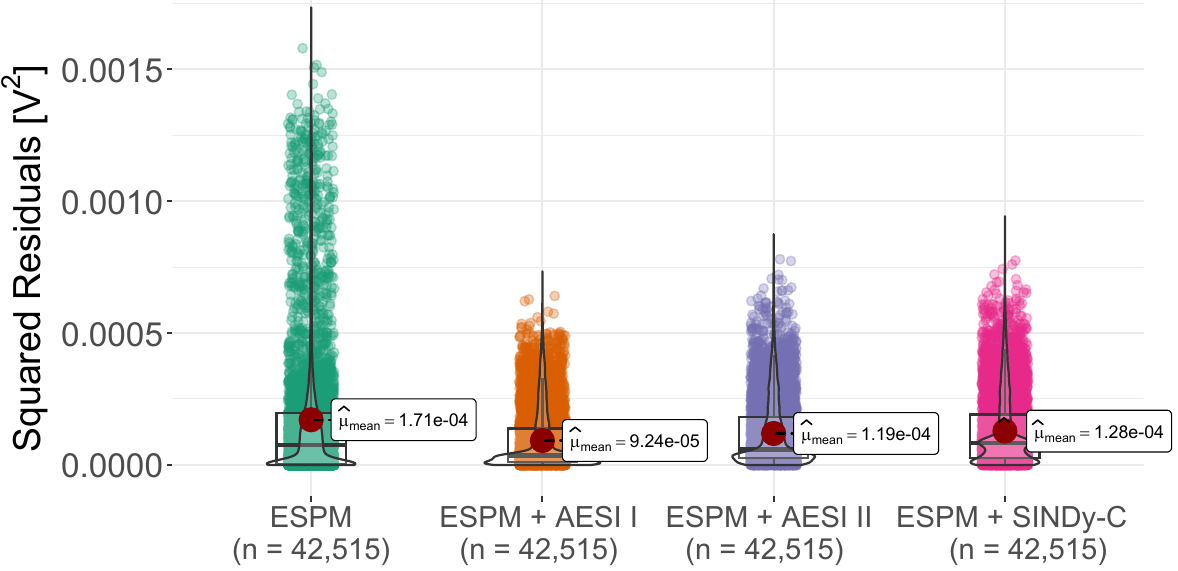}
    \caption{Comparative analysis between PBM and hybrid models}
    \label{fig:violin_plot}
    \end{center}
\end{figure}

Figure~\ref{fig:plot_actual_hybrd} illustrates the voltage predictions across the training, validation, and testing sets, comparing the ESPM, the ESPM + AESI I hybrid model, and experimental measurements. The results show that the hybrid model significantly reduces the discrepancy between predicted and measured voltages, particularly during the testing phase. Further insights are provided in Fig.~\ref{fig:violin_plot}, which shows the distribution of squared residuals for all models. The residuals from the hybrid models are more tightly concentrated around zero compared to those from the ESPM alone, indicating a substantial reduction in both average error and variability. This reflects not only improved accuracy but also enhanced consistency, as evidenced by lower MSER values and a narrower interquartile range.

\subsection{Analysis of the Basis Functions in the AESI}

\subsubsection{Feature Importance Ranking} In this study, feature importance is assessed using Singular Value Decomposition (SVD), which factorizes the rectangular matrix $\mathcal{S}$ into three components:
\begin{equation} \label{eq:28}
\mathcal{S}_{(n \times m)} = U \Sigma V^{*}
\end{equation}
where $\mathcal{S}$ represents the feature matrix of the model, with each column corresponding to a specific predictor evaluated over time:
\begin{equation} \label{eq:29}
\mathcal{S}_{(n \times m)} = \begin{bmatrix}
    \Theta_{1,1}\xi_{1} & \cdots & \Theta_{1,m}\xi_{m} \\
    \vdots & \ddots & \vdots \\
     \Theta_{n,1}\xi_{1} & \cdots & \Theta_{n,m}\xi_{m}
\end{bmatrix}
\end{equation}

Each column of $\mathcal{S}$ can be approximated as a linear combination of the left singular vectors $U$:
\begin{equation} \label{eq:30}
\begin{aligned}
    \mathcal{S}_{1} &= u_{1,1} \overrightarrow{U_1} + u_{1,2} \overrightarrow{U_2} + \cdots + u_{1,m} \overrightarrow{U_{m}} \\
    & \vdots \\
    \mathcal{S}_{m} &= u_{m,1} \overrightarrow{U_1} + u_{m,2} \overrightarrow{U_2} + \cdots + u_{m,m} \overrightarrow{U_{m}}
\end{aligned}
\end{equation}
The coefficients $u_{i,k}$, which associate individual predictors with the basis vectors in $U$, are determined by solving the following least squares problem:
\begin{equation}  \label{eq:32}
\tilde{\mathbf{u}}_j = \arg\min_{\mathbf{u}_j \in \mathbb{R}^{m}} \| \mathcal{S}_j - \mathbf{u}_j \mathbf{U} \|_2^2, \quad \forall ~ j = 1, \ldots, m
\end{equation}
A feature importance score $\bar{x}i$ is then computed as the weighted average of the coefficients $u{i,k}$, using the singular values $\sigma_k$ as weights:
\begin{equation}
\bar{x}_{i} = \frac{\sum_{k=1}^{m} u_{i,k} \sigma_k}{\sum_{k=1}^{n} \sigma_k}, \quad \forall ~ i = 1, \ldots, m
\end{equation}
\subsubsection{Model structure analysis} Assigning physical meaning to individual features is challenging, as it requires ground-truth measurements for internal battery states that are experimentally inaccessible. Moreover, voltage error correction is not uniquely defined, as multiple analytical expressions may sufficiently capture the nonlinear system dynamics. The lack of explicit physical constraints in the optimization process further complicates interpretability.

\begin{figure}[h!] 
    \begin{center}
       \includegraphics[angle=0,scale=0.41]{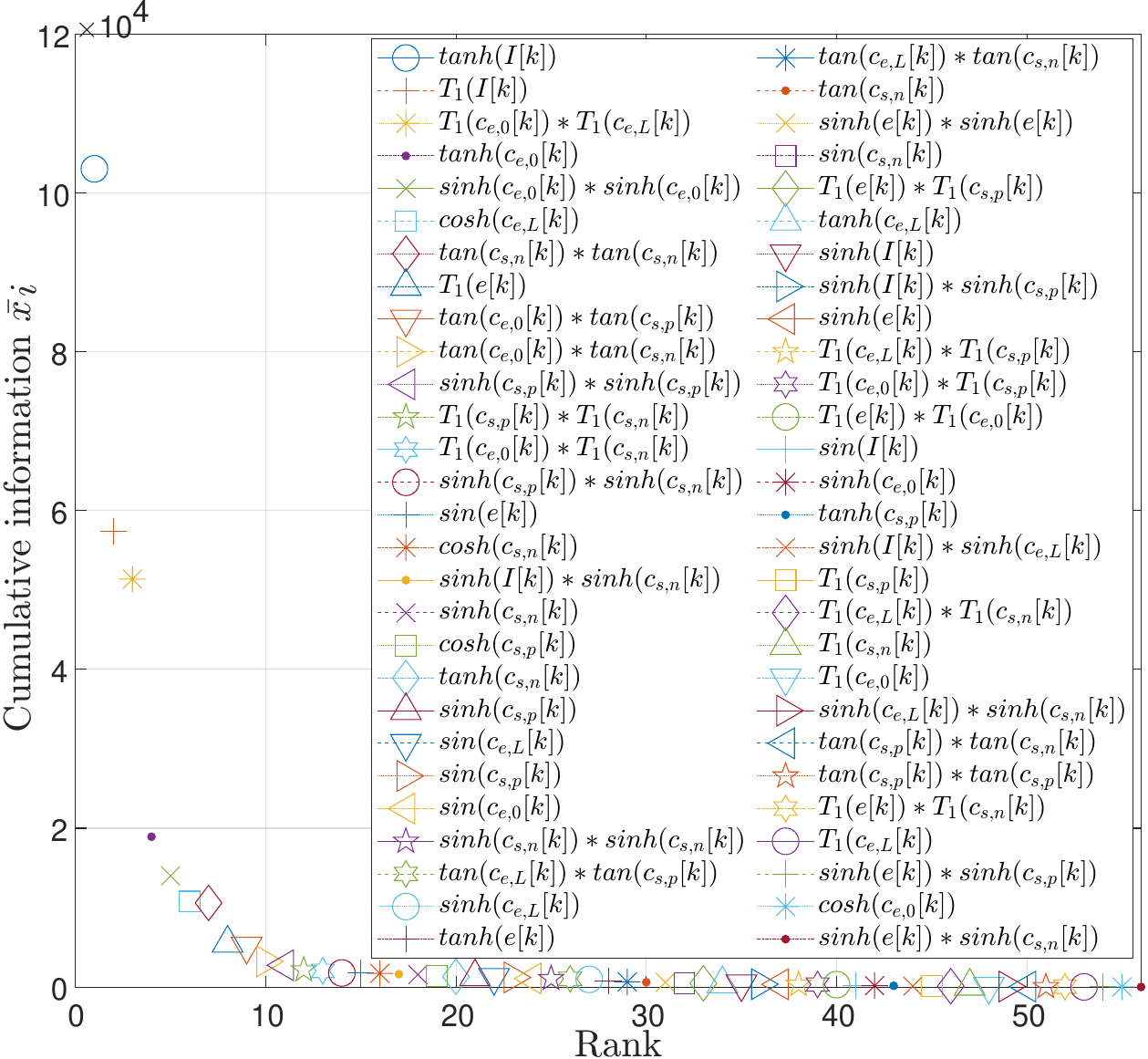}
    \caption{SVD ranking of the discovered features from AESI I (with no intercept)}
    \label{fig:SVD1}
    \end{center}
\end{figure}

Nevertheless, insights can be drawn by analyzing the relative importance of features in the AESI models, as revealed by the SVD rankings. As shown in Fig. \ref{fig:SVD1}, the AESI I model presents feature combinations resulting in an aggregated sparse model containing 56 active basis functions. The complexity reduction relative to the initial library matrix $\Theta$ (Eq. \ref{eq:theta_split}) is significant, as it originally comprises 626 possible feature combinations.

Notably, the current $I[k]$ emerges as an important predictor, reflected by highly ranked terms such as $T_1(I[k])$ and $tanh(I[k])$, indicating that voltage error corrections are strongly influenced by current-dependent dynamics. This finding aligns with Fig.~\ref{fig:error_map_2}, where the largest errors in ESPM predictions occur at extreme current values (e.g., $\pm$100 A). 

The inclusion of cross-terms involving boundary electrolyte concentrations (e.g.,  ($T_1(c_{e,0}[k])T_1(c_{e,L}[k])$) suggests the model captures electrolyte transport effects not fully represented by the ESPM. As discussed in Section~\ref{section_2}, these concentration profiles are approximated via the Galerkin projection method \cite{8008775}). Electrode solid-phase concentrations appear frequently in nonlinear trigonometric and hyperbolic terms, indicating the model's sensitivity to diffusion-limited dynamics within the electrodes. Prior voltage error terms $e[k]$ also contribute, emphasizing the temporal correlation and memory effects present in voltage error evolution.

\subsection{Analysis of the Uncertainty Quantification in AESI}
\begin{figure}[!h]
\centering
\subfigure[Testing set]{
    \includegraphics[width=0.41\textwidth]{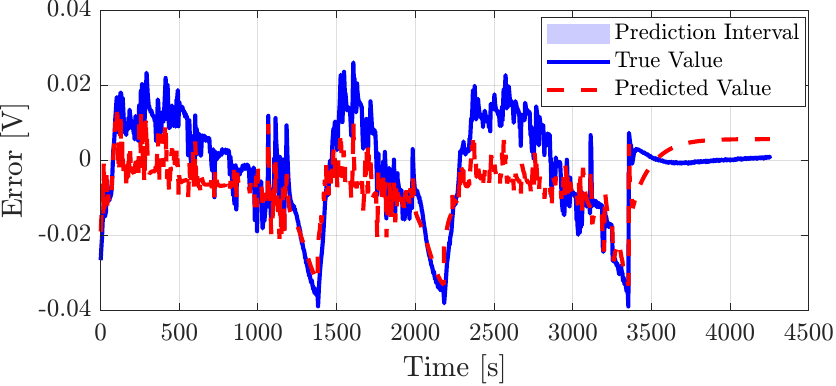}
}
\subfigure[Zoomed plot]{
    \includegraphics[width=0.395\textwidth]{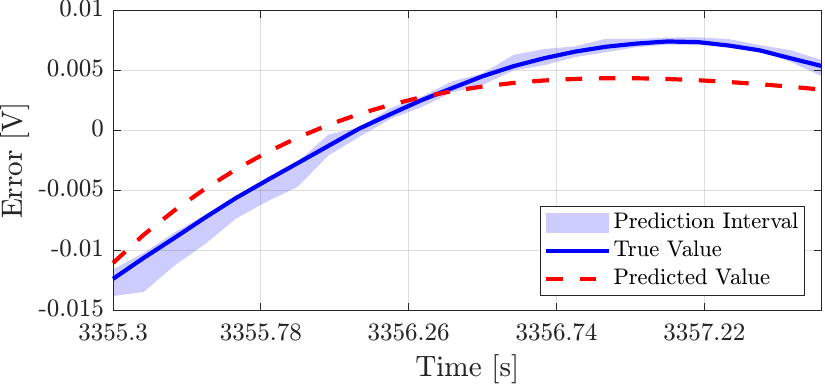}
}
\caption{SPCI uncertainty quantification of AESI I in testing.}
\label{fig:error_AESI_I}
\end{figure}

\begin{figure}[!h]
\centering
\subfigure[Testing set]{
    \includegraphics[width=0.41\textwidth]{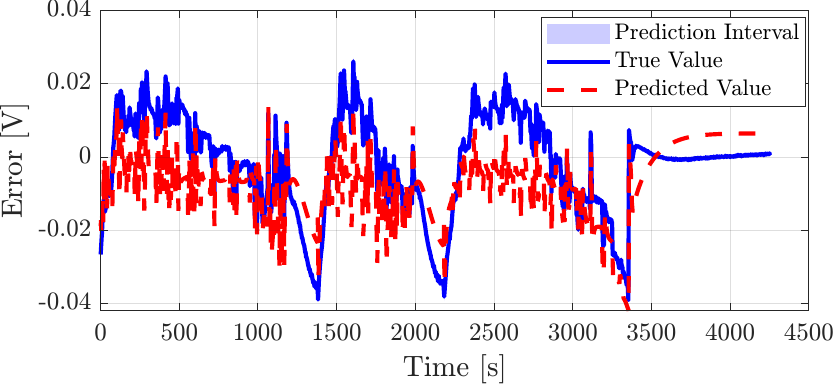}
}
\subfigure[Zoomed plot]{
    \includegraphics[width=0.395\textwidth]{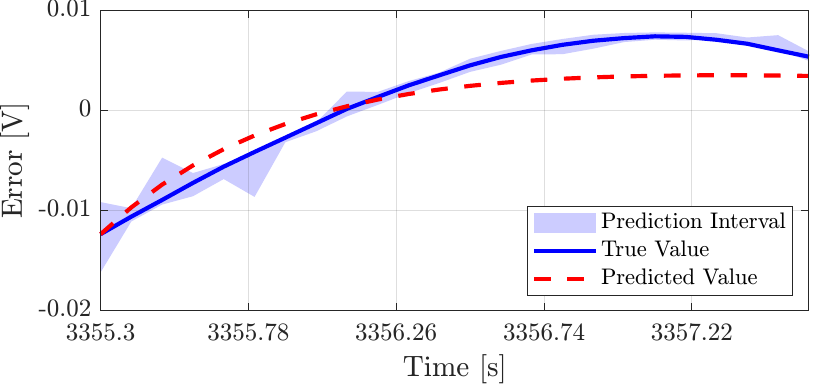}
}
\caption{SPCI uncertainty quantification of AESI II 
in testing.}
\label{fig:error_AESI_II}
\end{figure}

To evaluate uncertainty quantification in the AESI framework, Fig.~\ref{fig:error_AESI_I} and Fig.~\ref{fig:error_AESI_II} illustrate the performance of the SPCI method when applied to AESI-I and AESI-II, respectively. Each figure shows the ground truth voltage error, predicted values, and associated conformal prediction intervals (shaded regions) over the testing dataset. The prediction intervals dynamically adapt to the temporal variability of the residuals, consistently enclosing the true values. Zoomed-in views in the subplots further highlight the model’s ability to maintain tight yet valid prediction intervals, even during transient dynamics.

These visual observations are supported by the quantitative results summarized in Table~\ref{table:SPCI}. The SPCI method achieves empirical coverage rates of 96.85\% and 97.41\% for AESI-I and AESI-II, respectively—well above the nominal target of 90\%. Crucially, this high coverage is attained with relatively narrow average prediction intervals, indicating an effective balance between reliability and informativeness.

Overall, these results confirm that the SPCI framework satisfies the validity criteria of conformal prediction while offering efficient and interpretable uncertainty quantification for hybrid battery models.

%For the uncertainty quantification analysis of the AESI, the ground truth, predicted values using AESI I and II and prediction intervals constructed by the SPCI for the testing data are plotted in Fig. \ref{fig:error_AESI_I} and Fig. \ref{fig:error_AESI_II}. Results show that the SPCI approach can construct prediction intervals that can successfully cover the ground truth trajectories. These findings are supported by the numerical results in Table \ref{table:SPCI}, where the coverage rate and average width of the computed prediction intervals of the SPCI approach are shown. It can be seen that the SPCI approach can surpass the target coverage of 90\%  with narrow prediction intervals when integrated with both AESI approaches, satisfying the validity of the conformal prediction as required. 
\begin{table}[h!]
\centering
\caption{Comparison of SCPI for ensemble models in testing}
\label{table:SPCI}
\begin{tabular}{lcc}
\hline
\textbf{Model} & \textbf{Coverage [\%]} & \textbf{Average Width} [V]\\ 
\hline
AESI I     & $96.85$ & 3.39e-4 \\
AESI II    &  $97.41$  & 4.36e-4\\
\hline
\end{tabular}
\end{table}

% \textcolor{red}{Physics-based model; SINDy using STRidge regression and grid search optimization for hyperparameter tuning - Pure data}

% For SINDy (pure data): 
% V[k+1] = f(V[k],SOC[k],I[k])

% \section{Algorithms}
% ...

% \begin{algorithm}[H]
% \caption{Weighted Tanimoto ELM.}\label{alg:alg1}
% \begin{algorithmic}
% \STATE 
% \STATE {\textsc{TRAIN}}$(\mathbf{X} \mathbf{T})$
% \STATE \hspace{0.5cm}$ \textbf{select randomly } W \subset \mathbf{X}  $
% \STATE \hspace{0.5cm}$ N_\mathbf{t} \gets | \{ i : \mathbf{t}_i = \mathbf{t} \} | $ \textbf{ for } $ \mathbf{t}= -1,+1 $
% \STATE \hspace{0.5cm}$ B_i \gets \sqrt{ \textsc{max}(N_{-1},N_{+1}) / N_{\mathbf{t}_i} } $ \textbf{ for } $ i = 1,...,N $
% \STATE \hspace{0.5cm}$ \hat{\mathbf{H}} \gets  B \cdot (\mathbf{X}^T\textbf{W})/( \mathbb{1}\mathbf{X} + \mathbb{1}\textbf{W} - \mathbf{X}^T\textbf{W} ) $
% \STATE \hspace{0.5cm}$ \beta \gets \left ( I/C + \hat{\mathbf{H}}^T\hat{\mathbf{H}} \right )^{-1}(\hat{\mathbf{H}}^T B\cdot \mathbf{T})  $
% \STATE \hspace{0.5cm}\textbf{return}  $\textbf{W},  \beta $
% \STATE 
% \STATE {\textsc{PREDICT}}$(\mathbf{X} )$
% \STATE \hspace{0.5cm}$ \mathbf{H} \gets  (\mathbf{X}^T\textbf{W} )/( \mathbb{1}\mathbf{X}  + \mathbb{1}\textbf{W}- \mathbf{X}^T\textbf{W}  ) $
% \STATE \hspace{0.5cm}\textbf{return}  $\textsc{sign}( \mathbf{H} \beta )$
% \end{algorithmic}
% \label{alg1}
% \end{algorithm}

\section{Conclusions and Future Works} \label{section_6}

This study introduces the AESI framework, which combines physics-based modeling with data-driven sparse learning to improve the accuracy and robustness of lithium-ion battery predictions. By augmenting the ESPM with ensembles of sparse regressors trained on MBB samples, the proposed hybrid model effectively captures dynamics missed by reduced-order models, especially under high C-rate conditions. 

The ensemble strategies, namely bagging and stability selection, were evaluated in this work. Both enhanced voltage prediction accuracy relative to the ESPM alone, with mean squared error reductions up to 46\% on unseen testing data. While the stability-based approach achieved the best fit on training data, the bagging-based model demonstrated superior generalization, indicating stronger robustness. Model uncertainty was quantified using SPCI, a conformal prediction method tailored for non-exchangeable residuals. It produced statistically valid prediction intervals with coverage rates of 96.85\% (bagging) and 97.41\% (stability), supporting the reliability of the hybrid framework.

Future efforts will explore online learning capabilities for real-time diagnostics and integration into control schemes such as Nonlinear Model Predictive Control (NMPC). Another direction involves extending AESI with degradation-aware features to track long-term SOH, incorporating mechanisms such as loss of lithium inventory (LLI) and active material (LAM). While degradation evolves more slowly than voltage dynamics, this can be addressed through multi-timescale modeling or by periodically updating the hybrid library with aging-informed features such as ampere-hour throughput and cycle count.
\section*{Acknowledgements}
The authors would like to express their gratitude to Dr. Phillip Aquino of the Honda Research Institute for the insightful discussions and valuable feedback that contributed to the development of this research.

\bibliographystyle{IEEEtran}
\bibliography{mybibfile}

\begin{thebibliography}{10}
\providecommand{\url}[1]{#1}
\csname url@samestyle\endcsname
\providecommand{\newblock}{\relax}
\providecommand{\bibinfo}[2]{#2}
\providecommand{\BIBentrySTDinterwordspacing}{\spaceskip=0pt\relax}
\providecommand{\BIBentryALTinterwordstretchfactor}{4}
\providecommand{\BIBentryALTinterwordspacing}{\spaceskip=\fontdimen2\font plus
\BIBentryALTinterwordstretchfactor\fontdimen3\font minus \fontdimen4\font\relax}
\providecommand{\BIBforeignlanguage}[2]{{%
\expandafter\ifx\csname l@#1\endcsname\relax
\typeout{** WARNING: IEEEtran.bst: No hyphenation pattern has been}%
\typeout{** loaded for the language `#1'. Using the pattern for}%
\typeout{** the default language instead.}%
\else
\language=\csname l@#1\endcsname
\fi
#2}}
\providecommand{\BIBdecl}{\relax}
\BIBdecl

\bibitem{cano2018batteries}
Z.~P. Cano, D.~Banham, S.~Ye, A.~Hintennach, J.~Lu, M.~Fowler, and Z.~Chen, ``Batteries and fuel cells for emerging electric vehicle markets,'' \emph{Nature Energy}, vol.~3, no.~4, pp. 279--289, 2018.

\bibitem{CHEN20194363}
W.~Chen, J.~Liang, Z.~Yang, and G.~Li, ``A review of lithium-ion battery for electric vehicle applications and beyond,'' \emph{Energy Procedia}, vol. 158, pp. 4363--4368, 2019, innovative Solutions for Energy Transitions.

\bibitem{ALI2024144360}
H.~A.~A. Ali, L.~H. Raijmakers, K.~Chayambuka, D.~L. Danilov, P.~H. Notten, and R.-A. Eichel, ``A comparison between physics-based li-ion battery models,'' \emph{Electrochimica Acta}, vol. 493, p. 144360, 2024.

\bibitem{YU2023107661}
H.~Yu, L.~Zhang, W.~Wang, K.~Yang, Z.~Zhang, X.~Liang, S.~Chen, S.~Yang, J.~Li, and X.~Liu, ``Lithium-ion battery multi-scale modeling coupled with simplified electrochemical model and kinetic monte carlo model,'' \emph{iScience}, vol.~26, no.~9, p. 107661, 2023.

\bibitem{fan2016modeling}
G.~Fan, K.~Pan, M.~Canova, J.~Marcicki, and X.~G. Yang, ``Modeling of li-ion cells for fast simulation of high c-rate and low temperature operations,'' \emph{Journal of The Electrochemical Society}, vol. 163, no.~5, p. A666, 2016.

\bibitem{lai2021new}
Q.~Lai, H.~J. Ahn, Y.~Kim, Y.~N. Kim, and X.~Lin, ``New data optimization framework for parameter estimation under uncertainties with application to lithium-ion battery,'' \emph{Applied Energy}, vol. 295, p. 117034, 2021.

\bibitem{huang2023reinforcement}
R.~Huang, J.~Fogelquist, and X.~Lin, ``Reinforcement learning of optimal input excitation for parameter estimation with application to li-ion battery,'' \emph{IEEE Transactions on Industrial Informatics}, vol.~19, no.~11, pp. 11\,160--11\,170, 2023.

\bibitem{mikesell2025real}
I.~Mikesell, S.~F. da~Silva, M.~F. Ozkan, F.~E. Idrissi, P.~Ramesh, and M.~Canova, ``Real-time optimal design of experiment for parameter identification of li-ion cell electrochemical model,'' \emph{arXiv preprint arXiv:2504.15578}, 2025.

\bibitem{liang2022comparative}
Y.~Liang, A.~Emadi, O.~Gross, C.~Vidal, M.~Canova, S.~Panchal, P.~Kollmeyer, M.~Naguib, and F.~Khanum, ``A comparative study between physics, electrical and data driven lithium-ion battery voltage modeling approaches,'' SAE Technical Paper, Tech. Rep., 2022.

\bibitem{miranda2023particle}
M.~H. Miranda, F.~L. Silva, M.~A. Louren{\c{c}}o, J.~J. Eckert, and L.~C. Silva, ``Particle swarm optimization of elman neural network applied to battery state of charge and state of health estimation,'' \emph{Energy}, vol. 285, p. 129503, 2023.

\bibitem{amiri2024lithium}
M.~N. Amiri, A.~H{\aa}kansson, O.~S. Burheim, and J.~J. Lamb, ``Lithium-ion battery digitalization: Combining physics-based models and machine learning,'' \emph{Renewable and Sustainable Energy Reviews}, vol. 200, p. 114577, 2024.

\bibitem{pozzato2023combining}
G.~Pozzato and S.~Onori, ``Combining physics-based and machine learning methods to accelerate innovation in sustainable transportation and beyond: a control perspective,'' in \emph{2023 American Control Conference (ACC)}.\hskip 1em plus 0.5em minus 0.4em\relax IEEE, 2023, pp. 640--653.

\bibitem{tu2023integrating}
H.~Tu, S.~Moura, Y.~Wang, and H.~Fang, ``Integrating physics-based modeling with machine learning for lithium-ion batteries,'' \emph{Applied energy}, vol. 329, p. 120289, 2023.

\bibitem{krewer2018dynamic}
U.~Krewer, F.~R{\"o}der, E.~Harinath, R.~D. Braatz, B.~Bed{\"u}rftig, and R.~Findeisen, ``Dynamic models of li-ion batteries for diagnosis and operation: a review and perspective,'' \emph{Journal of the electrochemical society}, vol. 165, no.~16, p. A3656, 2018.

\bibitem{de2020discovery}
B.~M. De~Silva, D.~M. Higdon, S.~L. Brunton, and J.~N. Kutz, ``Discovery of physics from data: Universal laws and discrepancies,'' \emph{Frontiers in artificial intelligence}, vol.~3, p.~25, 2020.

\bibitem{zhang2024discovering}
Z.~Zhang, Z.~Zou, E.~Kuhl, and G.~E. Karniadakis, ``Discovering a reaction--diffusion model for alzheimer’s disease by combining pinns with symbolic regression,'' \emph{Computer Methods in Applied Mechanics and Engineering}, vol. 419, p. 116647, 2024.

\bibitem{bhatt2023sindy}
N.~Bhatt, B.~Jayawardhana, and S.~S.-E. Plaza, ``Sindy-crn: Sparse identification of chemical reaction networks from data,'' in \emph{2023 62nd IEEE Conference on Decision and Control (CDC)}.\hskip 1em plus 0.5em minus 0.4em\relax IEEE, 2023, pp. 3512--3518.

\bibitem{AHMADZADEH2024110743}
O.~Ahmadzadeh, Y.~Wang, and D.~Soudbakhsh, ``A data-driven framework for learning governing equations of li-ion batteries and co-estimating voltage and state-of-charge,'' \emph{Journal of Energy Storage}, vol.~84, p. 110743, 2024.

\bibitem{choi2023data}
H.~Choi, V.~De~Angelis, and Y.~Preger, ``Data-driven battery modeling based on koopman operator approximation using neural network,'' in \emph{2023 IEEE Power \& Energy Society General Meeting (PESGM)}.\hskip 1em plus 0.5em minus 0.4em\relax IEEE, 2023, pp. 1--5.

\bibitem{flores2022learning}
E.~Flores, C.~W{\"o}lke, P.~Yan, M.~Winter, T.~Vegge, I.~Cekic-Laskovic, and A.~Bhowmik, ``Learning the laws of lithium-ion transport in electrolytes using symbolic regression,'' \emph{Digital Discovery}, vol.~1, no.~4, pp. 440--447, 2022.

\bibitem{keren2023computational}
L.~S. Keren, A.~Liberzon, and T.~Lazebnik, ``A computational framework for physics-informed symbolic regression with straightforward integration of domain knowledge,'' \emph{Scientific Reports}, vol.~13, no.~1, p. 1249, 2023.

\bibitem{da2024improving}
S.~F. da~Silva, M.~F. Ozkan, F.~E. Idrissi, P.~Ramesh, and M.~Canova, ``Improving low-fidelity models of li-ion batteries via hybrid sparse identification of nonlinear dynamics,'' \emph{arXiv preprint arXiv:2411.12935}, 2024.

\bibitem{sashidhar2022bagging}
D.~Sashidhar and J.~N. Kutz, ``Bagging, optimized dynamic mode decomposition for robust, stable forecasting with spatial and temporal uncertainty quantification,'' \emph{Philosophical Transactions of the Royal Society A}, vol. 380, no. 2229, p. 20210199, 2022.

\bibitem{lee2023battery}
K.-J. Lee, W.-H. Lee, and K.-K.~K. Kim, ``Battery state-of-charge estimation using data-driven gaussian process kalman filters,'' \emph{Journal of Energy Storage}, vol.~72, p. 108392, 2023.

\bibitem{yang2018novel}
D.~Yang, X.~Zhang, R.~Pan, Y.~Wang, and Z.~Chen, ``A novel gaussian process regression model for state-of-health estimation of lithium-ion battery using charging curve,'' \emph{Journal of Power Sources}, vol. 384, pp. 387--395, 2018.

\bibitem{vovkconformal}
V.~Vovk, A.~Gammerman, and G.~Shafer, \emph{Algorithmic learning in a random world}.\hskip 1em plus 0.5em minus 0.4em\relax Springer, 2005, vol.~29.

\bibitem{manokhin2023practical}
V.~Manokhin, \emph{Practical Guide to Applied Conformal Prediction in Python}.\hskip 1em plus 0.5em minus 0.4em\relax Packt Publishing Ltd., Birmingham, 2023.

\bibitem{8008775}
G.~Fan, X.~Li, and M.~Canova, ``A reduced-order electrochemical model of li-ion batteries for control and estimation applications,'' \emph{IEEE Transactions on Vehicular Technology}, vol.~67, no.~1, pp. 76--91, 2018.

\bibitem{brunton2016discovering}
S.~L. Brunton, J.~L. Proctor, and J.~N. Kutz, ``Discovering governing equations from data by sparse identification of nonlinear dynamical systems,'' \emph{Proceedings of the national academy of sciences}, vol. 113, no.~15, pp. 3932--3937, 2016.

\bibitem{quade2018sparse}
M.~Quade, M.~Abel, J.~Nathan~Kutz, and S.~L. Brunton, ``Sparse identification of nonlinear dynamics for rapid model recovery,'' \emph{Chaos: An Interdisciplinary Journal of Nonlinear Science}, vol.~28, no.~6, 2018.

\bibitem{mostajeran2025scaled}
F.~Mostajeran and S.~A. Faroughi, ``Scaled-cpikans: Domain scaling in chebyshev-based physics-informed kolmogorov-arnold networks,'' \emph{arXiv preprint arXiv:2501.02762}, 2025.

\bibitem{rudy2019data}
S.~Rudy, A.~Alla, S.~L. Brunton, and J.~N. Kutz, ``Data-driven identification of parametric partial differential equations,'' \emph{SIAM Journal on Applied Dynamical Systems}, vol.~18, no.~2, pp. 643--660, 2019.

\bibitem{lahiri2013resampling}
S.~N. Lahiri, \emph{Resampling methods for dependent data}.\hskip 1em plus 0.5em minus 0.4em\relax Springer Science \& Business Media, 2013.

\bibitem{fasel2022ensemble}
U.~Fasel, J.~N. Kutz, B.~W. Brunton, and S.~L. Brunton, ``Ensemble-sindy: Robust sparse model discovery in the low-data, high-noise limit, with active learning and control,'' \emph{Proceedings of the Royal Society A}, vol. 478, no. 2260, p. 20210904, 2022.

\bibitem{hastie2009elements}
T.~Hastie, R.~Tibshirani, J.~H. Friedman, and J.~H. Friedman, \emph{The elements of statistical learning: data mining, inference, and prediction}.\hskip 1em plus 0.5em minus 0.4em\relax Springer, 2009, vol.~2.

\bibitem{meinshausen2010stability}
N.~Meinshausen and P.~B{\"u}hlmann, ``Stability selection,'' \emph{Journal of the Royal Statistical Society Series B: Statistical Methodology}, vol.~72, no.~4, pp. 417--473, 2010.

\bibitem{splitconformal}
H.~Papadopoulos, V.~Vovk, and A.~Gammerman, ``Conformal prediction with neural networks,'' in \emph{19th IEEE International Conference on Tools with Artificial Intelligence (ICTAI 2007)}, vol.~2.\hskip 1em plus 0.5em minus 0.4em\relax IEEE, 2007, pp. 388--395.

\bibitem{SPCI-Hu}
C.~Xu and Y.~Xie, ``Sequential predictive conformal inference for time series,'' in \emph{Proceedings of the 40th International Conference on Machine Learning}, ser. Proceedings of Machine Learning Research, vol. 202.\hskip 1em plus 0.5em minus 0.4em\relax PMLR, 23--29 Jul 2023, pp. 38\,707--38\,727.

\bibitem{EnbPI-Hu}
------, ``Conformal prediction interval for dynamic time-series,'' in \emph{Proceedings of the 38th International Conference on Machine Learning}, ser. Proceedings of Machine Learning Research, vol. 139.\hskip 1em plus 0.5em minus 0.4em\relax PMLR, 18--24 Jul 2021, pp. 11\,559--11\,569.

\bibitem{qrf}
N.~Meinshausen and G.~Ridgeway, ``Quantile regression forests.'' \emph{Journal of machine learning research}, vol.~7, no.~6, 2006.

\end{thebibliography}

\newpage

%{\color{red}\section{Biography Section}
% your own custom macro containing the $\backslash${\tt{includegraphics}} command to make things
 

%\bf{If you include a photo:}\vspace{-33pt}
%\begin{IEEEbiography}[{\includegraphics[width=1in,height=1.25in,clip,keepaspectratio]{fig1}}]{Michael Shell}
%Use $\backslash${\tt{begin\{IEEEbiography\}}} and then for the 1st argument use $\backslash${\tt{includegraphics}} to declare and link the author photo.
%Use the author name as the 3rd argument followed by the biography text.
%\end{IEEEbiography}}

\vspace{11pt}

\vfill

\end{document}